\documentclass[fleqn,usenatbib]{mnras}

\usepackage{newtxtext,newtxmath}

\usepackage[T1]{fontenc}

\DeclareRobustCommand{\VAN}[3]{#2}
\let\VANthebibliography\thebibliography
\def\thebibliography{\DeclareRobustCommand{\VAN}[3]{##3}\VANthebibliography}

\usepackage{graphicx}	
\usepackage{amsmath}	

\title[\textit{PLATO}'s Observing Strategies]{Assessing \textit{PLATO}'s Discovery Potential across Different Observing Strategies}

\author[Y. N. E. Eschen et al. ]{
Yoshi Nike Emilia Eschen$^{1,2}$\thanks{E-mail: yoshi.eschen@warwick.ac.uk}, Michelle Kunimoto$^{3}$, Juan Cabrera$^{4}$, Adrien Leleu$^{5}$,  \newauthor Thomas G. Wilson$^{1}$, Heike Rauer$^{4,5}$, Alex Romanec$^{1}$,  François Bouchy$^{6}$,  Magali Deleuil$^{7}$, \newauthor Nicholas Jannsen$^{2}$, Don Pollacco$^{1}$, Toby Rodel$^{8}$,  Sophia Sulis$^{7}$, Stéphane Udry$^{6}$
\\
$^{1}$Department of Physics, University of Warwick, Gibbet Hill Road, Coventry CV4 7AL, United Kingdom\\
$^{2}$Isaac Newton Group of Telescopes, Apartado 321, E-38700 Santa Cruz de La Palma, Canary Islands, Spain\\
$^{3}$Department of Physics and Astronomy, University of British Columbia, 6224 Agricultural Road, Vancouver, BC V6T 1Z1, Canada\\
$^{4}$Institute of Space Research, German Aerospace Center (DLR), Rutherfordstrasse 2, 12489 Berlin, Germany\\
$^{5}$Institut fuer Geologische Wissenschaften, Freie Universitaet Berlin, Malteserstrasse 74-100,12249 Berlin, Germany\\
$^{6}$Observatoire astronomique de l'Université de Genève, Chemin Pegasi 51, 1290 Versoix, Switzerland\\
$^{7}$Aix Marseille Univ, CNRS, CNES, LAM, 38 rue Frédéric Joliot-Curie, 13388 Marseille, France \\
$^{8}$Astrophysics Research Centre, School of Mathematics and Physics, Queen’s University Belfast, Belfast, BT7 1NN, UK}
\date{Accepted XXX. Received YYY; in original form ZZZ}

\pubyear{\the\year{}}
\begin{document}
\label{firstpage}
\pagerange{\pageref{firstpage}--\pageref{lastpage}}
\maketitle

\begin{abstract}
The \textit{PLATO} mission, planned to launch in spring 2027, will observe a fixed Southern field, LOPS2, for at least 2 years with the goal of discovering and deriving precise parameters for planets as small as Earth around FGK dwarfs and sub-giants. Afterwards, \textit{PLATO} could continue to monitor the same field or move to a Northern field, LOPN1, overlapping with the \textit{Kepler} field. We assess these two observing strategies in terms of sensitivity of detecting small and long period planets, and subsequently we estimate yields for both observing strategies. Within this we account for including the \textit{Kepler} data in the LOPN1 field, monitoring of up to 8 years of the LOPS2 field, and the sensitivity improvement from combining \textit{TESS} and \textit{PLATO} data since both fields have been and will be continued to be observed with \textit{TESS}.

We find that by monitoring both fields, and hence more stars, the total number of newly discovered planets is maximised. However, to optimise the number of discovered small and long periods planets, long-term monitoring of one field produces the best results. 
The combination of \textit{Kepler} and \textit{PLATO} data improves detection sensitivities; however, as this only covers a fraction of the LOPN1 field and stellar sample, it improves the yields of small and long period planets only marginally. Hence, to detect and characterise small and long period planets in this sparsely explored exoplanet parameter space, a longer continuous observation of the LOPS2 field is best.
\end{abstract}




\section{Introduction}


For more than 30 years, the number of detected exoplanets has been growing. While the first exoplanets were detected with the radial velocity (RV) method \citep{Mayor_Queloz_51Peg_1995}, the majority of exoplanets to date have been found via the transit method. The main contribution of the more than 4000 exoplanets detected with the transit method was due to photometric space-missions. 

The first of these missions, \textit{CoRoT} \citep{baglin2006}, detected the first rocky planet outside the Solar System~\citep{leger2009,queloz2009}, but independently confirmed a limited number of planets~\citep{deleuil2018}. 
To date, the main contribution was NASA's \textit{Kepler} mission \citep{Borucki_Kepler_2010} that has detected 2787 planets \citep[][accessed on 21. August 2026]{Exoplanet_Archive}. This mission was repurposed into \textit{K2} \citep{Howell_K2_2014} detecting 549 planets \citep[][accessed on 21. August 2026]{Exoplanet_Archive}. While these missions were pointing towards specific fields and only covering a fraction of the sky, the current NASA mission \textit{TESS} \citep{Ricker_TESS_2015} is performing an all-sky survey. To cover the entire sky, each area (so-called "sector") is observed for 27.4\,d that creates a bias in detecting planets of short orbital periods. 
These missions and their discoveries advanced the field of exoplanets enabling detailed discovery and demographic studies. Among the outputs of the \textit{Kepler} mission and further supported by the \textit{TESS} mission, planet types that do not exist in our Solar System were discovered, namely Sub-Neptunes and Super-Earths \citep{Batalha_Kepler_Candidates_2013}. Occurrence rate studies revealed that these planet types are the most common types in the Milky Way \citep{Howard_Kepler_Planet_Occurrence_2012,Petigura_super_earths_plateau_2013}. 

Combining the photometric observations with RV measurements, planetary radii and masses are characterised, and their interiors can be modelled \citep[e.g.][]{Egger_TOI469_plaNETic_2024, Eschen_TOI2345_2025}. Furthermore, atmospheric studies can be performed \citep[e.g.][]{Cherubim_LHS1140_2026} enabling a deeper understanding of the planet, and overall its formation and evolution. 
While gas giants planets, especially on short orbital periods, can be detected and characterised more easily due to their larger signals in photometric and spectroscopic/RV data, the detection and characterisation of small planets remains challenging due to limited precision of instruments and the impact of stellar activity. Furthermore, discovering planets of long orbital periods requires long-term monitoring to detect their long periodic signals. These factors limit the study of planets similar to the ones in our Solar System through which we can gain a deeper understanding of formation, evolution, and habitability, and also whether our Solar System is unique. 
Hence, to detect and characterise small and long period planets, orbiting their star in the habitable zone, long continuous observations at a high precision are required. To achieve this goal, the upcoming ESA mission \textit{PLATO} will be launched \citep{Rauer_PLATO_2014,Rauer_PLATO_2025}. 

To detect planets of orbital periods as long as Earth around these Sun-like stars, \textit{PLATO} will observe a fixed field in the Southern hemisphere, named LOPS2, for at least two years \citep{Nascimbeni_LOPS2_2025}. This field overlaps with the Southern \textit{TESS} continuous viewing zone (CVZ), where targets receive up to one year of continuous data. After two years, \textit{PLATO} could continue its stare at the LOPS2 field or move to a field in the North, LOPN1, that overlaps with the \textit{Kepler} field as well as the Northern \textit{TESS} CVZ. Observing strategies can be assessed based on the planet detection sensitivity and planet yield estimations, as computed for \textit{TESS} in \citet{Huang_TESS_Yield_2018}, \citet{Barclay_TESS_Yield_2018} and \citet{Kunimoto_TESS_Yield_2022}. Since the \textit{PLATO} mission follows the goal of detecting planets out to Earth's orbit, assessing its observing strategy in terms of detecting these types of planets is crucial to fulfil the mission objectives. To date, \textit{PLATO}'s planet yield has been assessed for two years of observations in the Southern field, followed by two years in the Northern field in \citet{Matuszewski_PLATO_Yield_2023} and \citet{cabrera2026} as well as possible observing strategy of 3 years in the South, followed by 1 year in the North \citep{Heller_PLATO_Yield_2022}. 

Following the study of \citet{Eschen_PLATO_Sensitivities_2024}, that showed the stellar and planetary content of the \textit{PLATO} LOPS2 field, in addition to computing and comparing sensitivities, we expand these sensitivities evaluate the discovery potential of the different \textit{PLATO} observing strategies. 
While \citet{Eschen_PLATO_Sensitivities_2024} focus on the sensitivities of the P1 sample of the LOPS2 field, and compared its sensitivities from simulated 2-year \textit{PLATO} observations to the real \textit{TESS} data, we expand this study by analysing the sensitivities of the Prime Sample, in particular computing the impact of longer stares up to 8 years in the southern field as well as the impact of the inclusion of \textit{TESS} and \textit{Kepler} data on the sensitivities and yields.

We introduce the missions of our analysis in \autoref{sec:missions}, outline our methods including the sample selection as well as sensitivity and yield computations in \autoref{sec:methods}, present our sensitivity and yields results in \autoref{sec:results} and finally discuss these in terms of \textit{PLATO}'s planet characterising goals accounting for stellar activity, ground-based follow-up, transit timing variation and the implications on occurrence rates in \autoref{sec:discussion}. We conclude in \autoref{sec:conclusions}.

\section{Space Missions Searching for Transiting Exoplanets}
\label{sec:missions}
Space-based missions have significantly contributed to the number of transiting exoplanets found to date. In the following, we provide an overview of the main space-based missions that have collected data on the targets in \textit{PLATO}'s Northern and Southern fields, \textit{Kepler} and \textit{TESS}. Each mission has a different set-up of the telescope, field of view, and observing strategy. With this, these missions have different goals they aim(ed) to reach and produce different data products.

\subsection{The \textit{Kepler} mission}
Significant contributions to the number of detected transiting exoplanets was brought by NASA's \textit{Kepler} mission. This mission was launched in 2009 and observed a fixed field of 115\,deg$^2$ for four years until 2013 split into observing quarters of $\sim$90\,d \citep{Borucki_Kepler_2010}. The \textit{Kepler} mission consisted of one telescope with a diameter of 0.95\,m and 42 CCDs. Individual images were processed into short- and long-cadence PDC-MAP lightcurves (Presearch Data Conditioning Maximum A Posteriori) \citep{Stumpe_PDCSAP_2012,Smith_PDCSAP_Kepler_2012} of $\sim$1 and $\sim$30\,min, respectively. The 48 months of {\it Kepler} data were split into Q0-Q17 observations of $\sim$200,000 stars with $\sim$112,000 objects observed in every quarter and $\sim$87,000 targets having data from a subset of quarters \citep{Seader2015} reported in the \textit{Kepler} Input Catalogue \citep[KIC;][]{Brown2011}. As only 512 stars were observed in short-cadence mode, we utilised the long-cadence data for consistency over a larger stellar sample. All light curves produced with the most recent DR24 \citep{Coughlin_Robovetter_2016} are available publicly via the Mikulski Archive for Space Telescopes (MAST).

\subsection{The Transiting Exoplanet Survey Satellite (\textit{TESS})}
In 2018 NASA launched the Transiting Exoplanet Survey Satellite \citep[\textit{TESS};][]{Ricker_TESS_2015}. \textit{TESS} consists of four cameras with a diameter of 10\,cm, each equipped with 4 CCDs. The combined field of view of the cameras sums up to 2300\,deg$^2$ and makes up one \textit{TESS} sector. The mission divided the entire sky into sectors in order to perform an all-sky survey over its mission duration. In doing so, a sector gets observed for 27.4\,days before the space-craft points to the next sector. The sectors overlap at the poles, resulting in stars in the CVZs being monitored continuously for up to a year per hemisphere monitoring respectively. 
This observing strategy results in a bias of detecting planets of short orbital periods \citep{Osborn_TOI2076_2022}. 
Targets that will be covered by \textit{TESS} are summarised in the \textit{TESS} Input Catalogue \citep[TIC;][]{Stassun_TIC_2018, Stassun_TIC_2019} containing roughly 1.5 billion sources over the entire sky.

\textit{TESS}'s observations are downlinked as Full Frame Images (FFIs) as well as target pixel files for selected targets.  These are processed by multiple pipelines, including the Science Processing Operations Center \citep[SPOC;][]{Jenkins_TESS_SPOC_2016, Caldwell_TESS_SPOC_2020} and the Quick Look Pipeline \citep[QLP;][]{Huang_QLP1_2020,Huang_QLP2_2020,Kunimoto_QLP_2021,Kunimoto_QLP_2022}. The FFIs, produced lightcurves and target pixel files are publicly available on MAST. Currently \textit{TESS} is in its third mission extension. During its Primary mission, the FFIs were binned at 30\,min cadence, in the first extension at 10\,min and since then FFIs are binned to 200\,s. Furthermore selected targets are observed at shorter cadences of 20\,s and 120\,s.

To date, the \textit{TESS} mission has found 929 confirmed planets as well as 5394  \textit{TESS} Objects of Interest (TOIs) that are being followed up \citep[][accessed on 21 August 2026]{Exoplanet_Archive,ExoFOP}.

\subsection{PLAnetary Transits and Oscillations of stars (\textit{PLATO})}
\begin{figure}
    \centering
    \includegraphics[width=\linewidth]{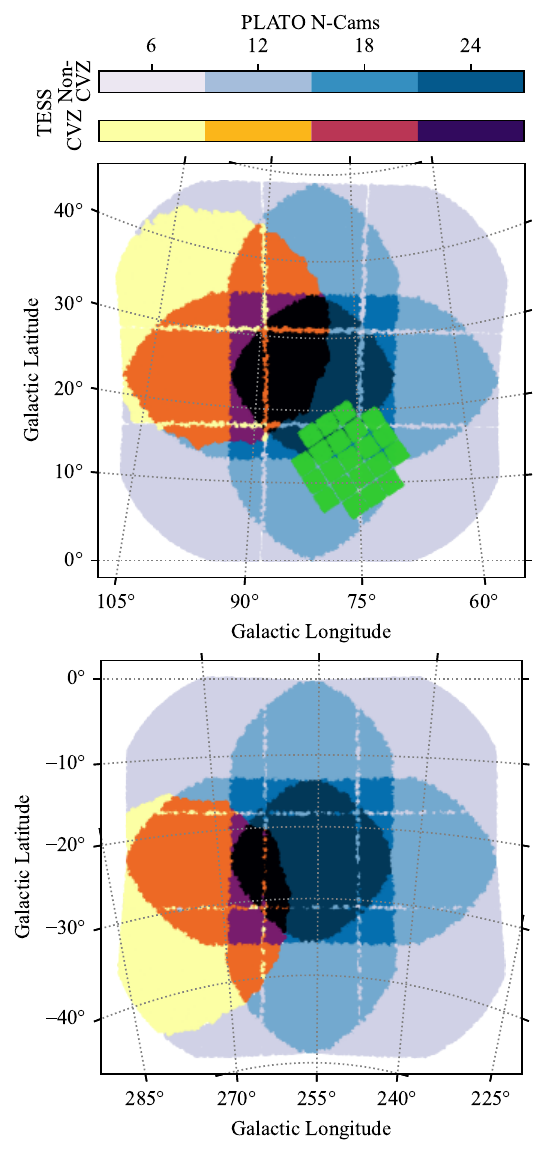}\vspace{-0.5cm}
    \caption{\textit{PLATO}'s observing fields in the North (top) and South (bottom) colour-coded by the number of N-Cams. Stars of the \textit{TESS} continuous viewing zones are respectively highlighted. The overlapping \textit{Kepler} field in the Northern field is shown in green.}
    \label{fig:plato_fields}
\end{figure}

To detect planets of longer orbital periods in the habitable zones of FGK stars, ESA will launch the PLAnetary Transits and Oscillations of stars \citep[\textit{PLATO};][]{Rauer_PLATO_2025} in early 2027. \textit{PLATO} is made up of 26 cameras, each with an entrance pupil of 12 cm and 4 CCDs per camera \citep[e.g.][]{cabrera2026}. While two of \textit{PLATO}'s cameras are F-CAMs which have a bluer and redder bandpass respectively, the remaining 24 cameras are N-CAMs with a wider bandpass and are grouped in 4 groups of 6. \textit{PLATO} will observe a fixed field in the South, LOPS2 \citep[][]{Nascimbeni_LOPS2_2022, Nascimbeni_LOPS2_2025}, for at least the first two years of its mission. This field covers $\sim$5\% of the sky. The N-CAMs overlap resulting in stars in the centre being observed with 24 cameras, while stars at the edges of the field get observed with 6 cameras as shown in \autoref{fig:plato_fields}. 

The stars \textit{PLATO} will observe are recorded in the \textit{PLATO} Input Catalogue of targets \citep[tPIC;][]{montalto2026}. Depending on stellar properties, the tPIC is divided into sub-samples:  P1, P2, P4 and P5. P1 (V<11\,mag) and P2 (V<8.5\,mag) contain the brightest FGK dwarfs and sub-giants of the PIC with the lowest predicted noise (<50\,ppm/h); P4 contains M dwarfs as faint as V=16\,mag and P5 is a statistical sample consisting of FGK dwarfs and sub-giants brighter than V=13\,mag \citep[][]{Rauer_PLATO_2025}. Furthermore, the \textit{PLATO} mission has a Prime sample of reserved pre-selected stars that are the most promising to find Sun-Earth analogues \citep{Nascimbeni_Prime_2026}. These samples are observed at different cadences and result in different data products, e.g. the Prime sample is observed at a cadence of 25\,s and the data downlinked as imagettes.  

\subsection{Observational Overlap of these Missions}
\label{sec:mission_overlap}
The \textit{PLATO} mission will start its observations with the Southern LOPS2 field. As shown in \citet{Eschen_PLATO_Sensitivities_2024} and in \autoref{fig:plato_fields}, the LOPS2 field overlaps with the CVZ of \textit{TESS} and hence these stars have already been monitored for up to 4 years. We show the \textit{TESS} coverage of the \textit{PLATO} targets in \autoref{fig:tess_sectors}.

After the first 2 years of the mission, \textit{PLATO} could move to a Northern field, LOPN1 \citep{Nascimbeni_LOPS2_2022, Rauer_PLATO_2025}. This field overlaps with the \textit{TESS} CVZ, but more importantly also with the \textit{Kepler} field as shown in \autoref{fig:plato_fields}. 
Since the current {\it PLATO} long-term observing strategy is not yet decided we aim to assess the (dis)advantages for each scenario accounting for the available data form overlapping regions.

\section{Methods} 
\label{sec:methods}
To assess the (dis-)advantages of each observing strategy, we compute the planet detection sensitivities and from them we estimate the planet yield.
In order to assess the detection sensitivities of planets for the different observing cases, we compute the sensitivities of the same subsets of stars for different observing strategies and compare our sensitivities and yields. In the following we describe how we select the samples, the observing strategies, and the computation of the planet detection sensitivities and yields for these.

\subsection{\textit{PLATO}'s Potential Observing Strategies}
To date, the full observing strategy of \textit{PLATO} is not confirmed. Using the overlap with \textit{Kepler} and \textit{TESS} as discussed in \autoref{sec:mission_overlap} and the possibility of longer observations, we compare the effects on sensitivities and yields for each scenario. With this we aim to answer the following questions:


\textit{\textbf{Question 1}: What are the (dis-)advantages of PLATO moving the North or remaining in the South after 2 years?}

\textbf{Case 1}: As \textit{PLATO}'s nominal mission duration is 4 years, we compare the confirmed 2 years of observations of the LOPS2 field to observing the same field for 4 years. The \textit{PLATO} mission carries enough consumables for a 4 year mission extension and hence could observe the LOPS2 field for up to 8 years. Hence we also compare the 2 and 4 year sensitivities to the sensitivities for 8 years of observations.

\textbf{Case 2}: The main difference of moving to the Northern field, LOPN1, is the overlap with the \textit{Kepler} field and its data availability. We hence combine the available \textit{Kepler} data with the simulated \textit{PLATO} data of 2 years and compare it to the sensitivities of observing with only \textit{PLATO} for 2 years. \\

\textit{\textbf{Question 2}: What is the impact of TESS for the detectability of PLATO planets?}

Since the LOPS2 field, as well as the LOPN1 field, has been monitored by \textit{TESS} since 2018, we compare the impact on the sensitivities of including the \textit{TESS} data to the 2 years of simulated \textit{PLATO} data against not including the \textit{TESS} data. As both fields contain a similar distribution of stars, we perform this comparison only for the Southern LOPS2 field as we expect similar results for the Northern LOPN1 field. 

\subsection{Selecting and Constructing Samples}
The previous work of \citet{Eschen_PLATO_Sensitivities_2024} used the P1 sample to compare the sensitivities of \textit{TESS} and \textit{PLATO}. We expand on this study to use the stars of \textit{PLATO}'s newly-defined Prime sample as this sample contains the most promising stars to detect Sun-Earth analogues and hence the stars we want to assess to maximise the output of the observing strategy. 

The Prime sample contains the most promising targets to find Earth-Sun-analogue and defined for the LOPS2 field following \citet{Nascimbeni_Prime_2026}. It is defined to 15000 bright stars across spectral ranges and planet hosting stars. 
Within the Prime sample, we divide between dwarfs and sub-giants following the radius cut between dwarfs and sub-giants \citep[$R_\ast=1+2\cdot10^{-7}(\text{Teff} - 4000)^2$,][]{Nascimbeni_LOPS2_2022}. We differentiate between M, K, G and F dwarfs following the temperature boundaries of \citet{Pecaut_Teff_2013} outlined in \citet{Nascimbeni_LOPS2_2022}, and divide the sub-giants using a temperature cut of 5500\,K \citep{Nascimbeni_LOPS2_2022} into hot and cool sub-giants. We show the samples and stellar distributions in \autoref{tab:samples}. As we aim to assess the sensitivity of detecting small planets, we focus on the dwarfs sample and show the results of the sub-giant sample in the appendix. We use the Prime sample for our analysis of Case 1, as well as to create the sample for \textit{TESS}. Since we consider the Northern field for Case 2 which does not have a defined Prime sample, we create our own Prime-like sample. 

Following the methods from \citet{Nascimbeni_Prime_2026}, we compute the photometric detection metric, $\mathcal{M}$, and follow-up metric, $\mathcal{R}$ for the tPIC of the LOPN1 field. We apply a radius cut of R$_*<1.5$\,R$_\odot$ and magnitude cut of V$<14$\,mag. Finally, we include all stars of the P2 sample and P4 stars which have H$<9$\,mag in 2MASS, which we compute following Equation 24 of \citet{Nascimbeni_Prime_2026}. Applying the same cut-off of the follow-up metric of $\mathcal{R}=0.73861$ gives us a sample of 12559 stars for our self-defined Northern Prime-like sample. We cross-match these targets with Kepler using the Kepler-Gaia catalogue from the Kepler-Gaia Crossmatch \footnote{\url{https://github.com/megbedell/gaia-kepler.fun}}. Finally, we apply the radius-cut used for LOPS2 to differentiate between dwarfs and sub-giants. We note that these metrics and cut-offs are optimised for the Southern field and may be updated during the PLATO operations for future observing strategies. However, utilising this method we are able to produce representative samples that are robust for our comparative exoplanet detection sensitivity analyses.

\begin{figure*}
    \centering
    \includegraphics[width=\linewidth]{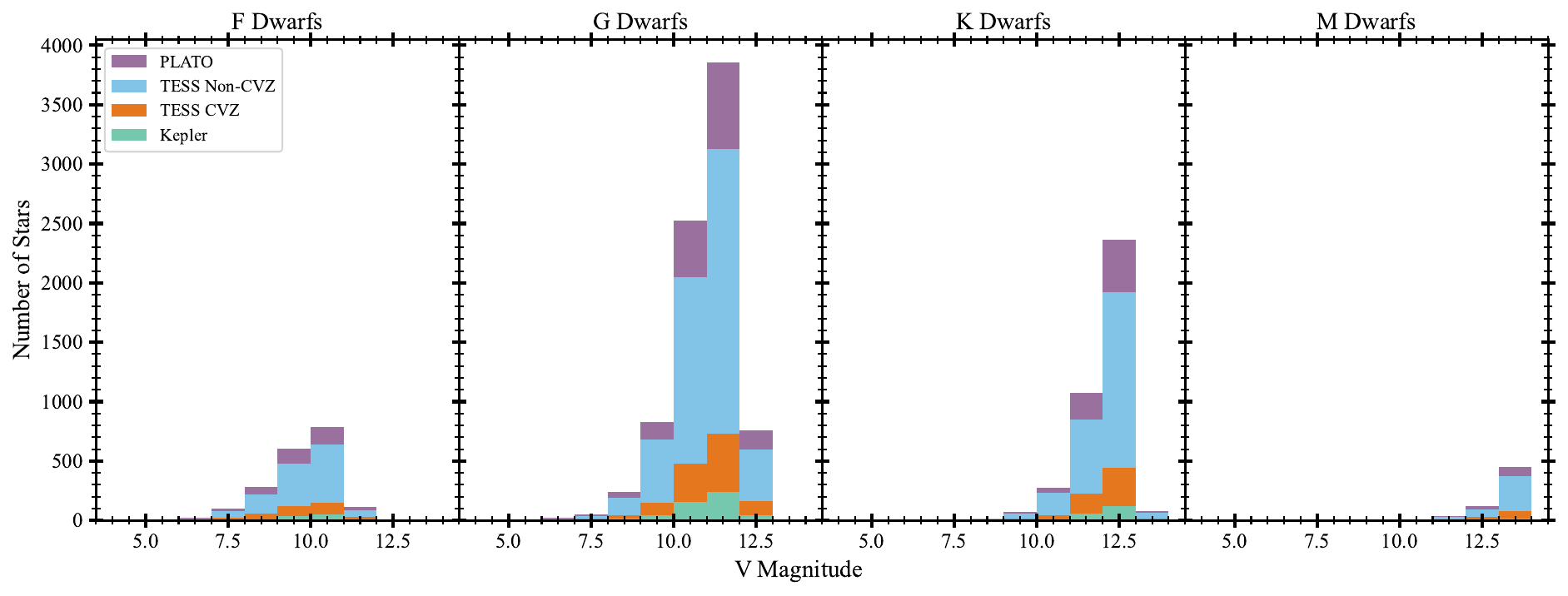}\vspace{-0.5cm}
    \caption{Histogram of magnitude of the dwarfs covered by \textit{PLATO} (purple), \textit{TESS} in the non-CVZ (blue) and CVZ (orange) in the LOPS2 field, and \textit{Kepler} (green) in the LOPN1 field for our Prime(-like) samples. We show the V magnitudes from the PIC for the F, G, K, and M Dwarfs separately.} 
    \label{fig:targets_magnitude_dwarfs}
\end{figure*}

\begin{figure}
    \centering
    \includegraphics[width=\linewidth]{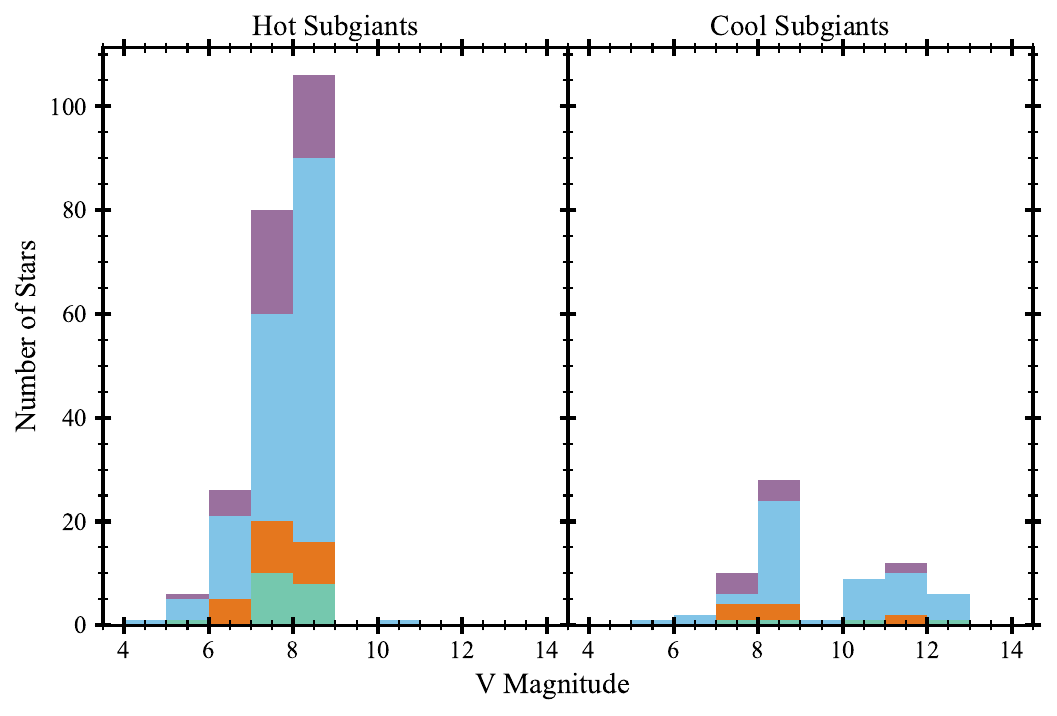}\vspace{-0.5cm}
    \caption{Same as \autoref{fig:targets_magnitude_dwarfs}, but for subgiants}
    \label{fig:targets_magnitude_subgiants}
\end{figure}

We show the magnitude distribution of the respective samples in \autoref{fig:targets_magnitude_dwarfs} and \autoref{fig:targets_magnitude_subgiants} for the different stellar types and summarise their sizes in \autoref{tab:samples}. 

To compute and compare overall yields, we use the entire tPIC. In the LOPS2 field, we use the Southern tPIC \citep{montalto2026} and filter for a number of N-CAMs above 0 to ensure the targets are observed. This results in 218146 targets for the South. In the Northern field, we cross-match the Northern PIC with the Kepler-Gaia catalogue as described above, which results in 13806 targets for our analysis of case 2. 

\begin{table*}
    \centering
    \caption{Sample sizes of the different stellar types we consider for the sensitivity computations of \textit{PLATO}'s Prime(-like) sample and its overlap with \textit{TESS} and \textit{Kepler} based on stellar types. Additionally, we show the total stars in the (t)PIC used for the yield computations.}
    \begin{tabular}{l|cccccccc}
    \hline
    \hline
    Sample & F Dwarfs & G Dwarfs & K Dwarfs & M Dwarfs & Hot Subgiants & Cool Subgiants & Total Prime(-like) & Total (t)PIC\\
    \hline
    \textit{PLATO} LOPS2 & 1917 & 8289 & 3883 & 619 & 220 & 69 & 14997 & 182581\\
    \textit{TESS} CVZ LOPS2 & 384 & 1587 & 743 & 110 & 42 & 10 & 2876 & 30441\\
    \textit{TESS} Non-CVZ LOPS2 & 1533 & 6702 & 3140 & 509 & 178 & 59 & 12121 & 152140\\
    \hline
    \textit{PLATO} LOPN1 & 1428 & 6862 & 3422 & 574 & 216 & 57 & 12559 & 175597\\
    \textit{Kepler} LOPN1 & 109 & 499 & 190 & 23 & 19 & 4 & 844 & 13806\\
    \hline
    \hline
    \end{tabular}
    \label{tab:samples}
\end{table*}

\subsection{Computing Sensitivities}
Following \citet{Eschen_PLATO_Sensitivities_2024}, we adapt the Transit Investigation and Recoverability Application \citep[\textsc{tiara};][]{Rodel_TIARA_2024}, which compares the signal-to-noise ratio of the depth of exoplanet transits to the noise in the photometric light curve data. This version, the \textit{\textbf{P}LATO} \textbf{E}xoplanet \textbf{S}ensitivity \textbf{To}ol (\textsc{pesto}), builds upon previous work by also computing averaged sensitivities for many observing cases. This provides the flexibility to use observed \textit{Kepler} and simulated \textit{PLATO} data in addition to \textit{TESS} data, and simulates planets robustly in a log$_2$ space ensuring all radii and periods are unbiasedly covered. 
We describe the full features of \textsc{pesto} in the following paragraphs.

\subsubsection{The \textit{\textbf{P}LATO} \textbf{E}xoplanet \textbf{S}ensitivity \textbf{To}ol (\textsc{pesto})}
The original \textsc{tiara} software was initially built for \textit{TESS} data and has been previously used for a {\it TESS} yield study \cite{Rodel_TIARA_2024}. It was modified to estimate planet detection sensitivities for the upcoming \textit{PLATO} mission \citep[][]{Eschen_PLATO_Sensitivities_2024}, whilst computing the detection signal-to-noise of the generated planets. However, as this version has been modified for the use of \textit{PLATO}, our input allows data from other missions, including simulated \textit{PLATO} data, to produce robust sensitivity outputs. This new development is able to conduct such assessment as it has been generalised to not depend on occurrence rates planet radius-orbital period bins. Since the \textsc{tiara} application simulated planets at radii and periods based on occurrence rates, it is possible that no or very few planets are simulated in given radius-period bins. To avoid this and obtain the sensitivities for any possible transiting planet, in our new implementation we inject 2000 planets randomly picked in period ranging from 0.5 to 1600\,d and radii from 0.3 to 16\,R$_\oplus$, uniformly distributed in log$_2$ space as done in \citet{Eschen_PLATO_Sensitivities_2024}. In order to fairly compare planet detection sensitivities across the radius-period parameter space, we set both the impact parameter (and with this the orbital inclination) and eccentricity to be 0. We acknowledge that an impact parameter of 0 is optimistic and hence we note that our sensitivities show an idealised scenario of detecting planets that transit perfectly edge-on.  

\textit{PLATO} will observe the stars of the Prime Sample with a cadence of 25\,seconds. However, to simulate data at this cadence for up to 8 years, requires a high computation time.
Hence, we bin the simulated timestamps into 10\,minutes as done in \citet{Eschen_PLATO_Sensitivities_2024}. However, this results in inaccurate sensitivities to planets at very short orbital periods and short transit durations as not enough data points fall into the transit window to flag the detection as significant. We account for detection sensitivities of these short orbital period planets by applying a smoothening function, which checks whether a planet of given radius could have been detected at a longer orbital period and then flags that this should have also been detected at a shorter orbital period. While this issue should not arise for planetary radii, we implement this functionality also in radius space, ensuring that if a planet at a given period could have been detected a larger planet at the same period should have also been detected with at least the same sensitivity. 

Finally, we add a feature to combine datasets and compute the sensitivity for observations with more than one telescope (e.g. {\em PLATO}, {\em Kepler}, and {\em TESS}). We describe this feature in more detail in \autoref{sec:combining_pesto}.
The sensitivity computations of \textsc{pesto} can be used to compute yields for single or combined missions as we describe in more detail in \autoref{sec:yields}.
Implementing all of these modifications, we introduce a modified version of \textsc{tiara} that has been optimised for the use of \textit{PLATO} and further comparison studies, \textsc{pesto}.

\subsubsection{\textsc{pesto}'s Applications to \textit{PLATO}, \textit{TESS}, and \textit{Kepler}}
\label{sec:pesto_data}
We create \textit{PLATO} timestamps by binning data to 10\,min. As \textit{PLATO} will rotate every 90\,days and downlink data, we account for data breaks of 1\,day in the rotation interval. We use the PIC stellar radius and target noise, computed by \textsc{pine} \citep[\texttt{BOLrandomSysNSR};][]{Boerner_PINE_2024}. These are the only inputs we require to compute the planet detection sensitivities using \textsc{pesto}.

In the same way, we compute these sensitivities to only analyse the \textit{Kepler} and \textit{TESS} data respectively. For \textit{Kepler} we use the observed timestamps and reported CDPP-noises as inputs to \textsc{pesto}. Since \textit{TESS} data is processed via different pipelines, we estimate the timestamps as done in \citet{Kunimoto_TESS_Yield_2022}. Using \textsc{tess-point}, we determine which targets in the PIC were observed in which \textit{TESS} sector on which CCD of which camera. For each possible combination of PIC targets, we randomly select one target, download its QLP lightcurve, filter for only timestamps of good quality (\texttt{QUALITY}==0) and use this as representative timestamps for all targets observed with the same combination of CCD, camera and sector.
As \textit{TESS} will continue to observe the LOPS2 field until sector 114, we query \textsc{tess-point} for sector and camera and simulate the timestamps for these upcoming sectors. For this, we create timestamps from the expected start and end date of each sector with a cadence of 200\,s. We account for downlink time by removing timestamps corresponding to 5 continuous hours weekly. For scattered light, we use the predicted degrees offset between each camera and the Moon and Earth. We remove points for each camera where this offset is below 35\,degrees. 

We estimate the 2-hour-noise as described in \autoref{app:tess_noise} to consider the 'best' case noise scenario for \textit{TESS} as the \textit{PLATO} noise values are. 

This noise estimation is dependent on the T magnitude which we obtain by following \citet{Stassun_TIC_2018} using the Ks and G magnitude where available and else the V and G magnitude. We use noise values and representative timestamps for all Prime targets in the respective sectors as inputs to \textsc{pesto} to compute the sensitivities for \textit{TESS}.

\subsubsection{Combining \textit{PLATO} data with \textit{TESS} and \textit{Kepler} in \textsc{pesto}}
\label{sec:combining_pesto}
As we also aim to assess the sensitivity when combining the predicted \textit{PLATO} data with the available data of the \textit{TESS} and \textit{Kepler} missions, we add another feature to \textsc{pesto} that enables us to combine datasets of different telescopes. For this, we combine the timestamps of \textit{TESS} or \textit{Kepler} with our simulated \textit{PLATO} timestamps ensuring that they are all in BJD and assuming a \textit{PLATO} launch date of 1 January 2027. Furthermore, we combine the noises of the respective missions matching the respective timestamps as input for \textsc{pesto}. We run \textsc{pesto} on the combined dataset. Since targets in the \textit{TESS} CVZ have significantly more \textit{TESS} data available, we differentiate between the \textit{TESS} CVZ and non-CVZ as done in \citet{Eschen_PLATO_Sensitivities_2024}.

To ensure a fair comparison between the samples created for the different cases, we simulate the same planets within each comparison. 

\subsection{Computing Yields with \textsc{pesto}}
\label{sec:yields}
In addition to the sensitivities, we implement yield calculations into \textsc{pesto}. While \textsc{tiara} multiplies occurrence rates with detection efficiencies and the number of stars, we follow the approach from \citet{Kunimoto_TESS_Yield_2022}. 
As our detections methods are biased towards detecting large and short period planets, occurrence rates do not span the entire grid of sensitivities we have computed. Hence, we first extrapolate occurrence rates up to 1600\,days and down to 0.3\,R$_\oplus$ as we have computed sensitivities for. We use the occurrences rates from \citet{Hsu_OccurrenceRates_2019} for FGK type stars and assume their bin densities are constant and account for their reported uncertainties. With this we extrapolate the occurrence rates for bins that only report an upper limit or are not considered (i.e. periods between 400 and 1600\,days and radii from 0.3-0.5\,R$_\oplus$). We first extrapolate horizontally in period. For each radius bin we use the longest period available and its bin density. In a second step, we extrapolate vertically in radii, for which we use the lowest radius available to extrapolate to smaller radii. For bins that report an upper limit on the occurrence rate, we check whether the extrapolated occurrence rate is below the upper limit. If this is the case, we use the extrapolated occurrence rate, else we set the occurrence rate of that bin to 0. We use the same methodology for M dwarfs for which we use the occurrence rates from \citet{Dressing_Charbonneau_Occurrence_2015}. Since these only span radii from 0.5 to 4\,R$_\oplus$, we extrapolate radii up to 16\,R$_\oplus$, for which we use the bin densities of the largest reported radii available for each period bin, as well as the smaller radii and longer orbital periods as we did for FGK type stars. We show our extrapolated occurrence rate grids used for this study in \autoref{fig:occurrence_rates_fgk} and \autoref{fig:occurrence_rates_m}.

We determine the number of planets around each star using the extrapolated occurrence rates and drawing from a Poisson distribution. We allocate each planet into a radius-period bin based on a probability obtained from the occurrence rate of the bin normalised by the overall occurrence rate for the respective spectral type. Within the bin, we obtain the radius and period for each planets by randomly drawing from a uniform distribution in logP and logR planet space, the cosine of the inclination from a uniform distribution between 0 and 1, the argument of periastron, $\omega$, from a uniform distribution between $-\pi$ and $+\pi$ and the eccentricity from a Beta distribution with $\alpha$ = 1.03 and $\beta$ = 13.6 \citep{VanEylen_Eccentricity_2015}. We use these to compute further planetary properties including the impact parameter. To ensure we only include transiting planets, we remove planets with an impact parameter above 1. For the remaining transiting planets, we compute the sensitivity by injecting 10 planets of the given planetary parameters at different transit epochs which are drawn from a uniform distribution in orbital phase. Using \textsc{pesto}'s sensitivity functionality as described before, we compute the expected signal-to-noise ratio and number of transits of these planets. As done in \citet{Kunimoto_TESS_Yield_2022}, we compute the detection probability using gamma functions. We use the gamma functions from \citet{Christiansen_KeplerPlanetDetection_2017} and \citet{Hsu_OccurrenceRates_2019}, which have been extrapolated to include duo- and mono-transits in \citet{Rodel_TIARA_2024}. We conduct a Bernoulli draw with the respective detection probability to determine whether a planet would be detected. As most transit searches require more than one transit to be detected, we report two results: one for the planet yield of planets with at least 3 observed transits, and the second one including mono-transits.
 We repeat this process 100 times allowing us to report the average values and uncertainties on detections.




\section{Results of each Observing Strategy}
\label{sec:results}
We present the result of each observing strategy as sensitivity maps divided into the spectral types (F, G, K and M dwarfs). Furthermore, to compare each scenario to two years of \textit{PLATO} observations we compute the sensitivity difference as done in \citet{Eschen_PLATO_Sensitivities_2024} and show these in comparison to the 2\,years of \textit{PLATO} of the respective sample (see \autoref{fig:PLATO_sensitivities}). The sensitivity ranges from yellow for a parameter space that is well detectable to dark purple which represents a parameter space that the underlying data and noise of the telescope(s) is/are not sensitive to. The sensitivity difference ranges from purple to orange, with purple highlighting areas where 2 years of \textit{PLATO} are more sensitive to than the observing strategy we compare to, and orange shows that this observing strategy results in a better detection sensitivity. In each plot we show the planets of the Solar System to highlight the space of Earth that we aim to characterise with the \textit{PLATO} mission.

We furthermore show the yields in radius-period grids for longer observations of the LOPS2 field, the inclusion of \textit{TESS} data, and for the combination of \textit{PLATO} and \textit{Kepler} in the LOPN1 field. As \textit{PLATO} has the key goal to detect small and long period planets, we compare the yields for specific thresholds in \autoref{tab:yields}. Within the table we show the total yields, the yields for planets of orbital periods beyond 100\,d, of radii below 1\,R$_\oplus$ and planets smaller than 1.5\,R$_\oplus$ in the habitable zone of its star which we computed following \citet{Kopparapu_HZ_Computation_2013}. We show these yields for requiring at least 3 transits as well as including mono- and duo-transits.

\begin{table*}
    \centering
    \caption{Computed yields for the LOPS2 field of 182581 stars for different observing strategies and the LOPN1 field of 175597 accounting for \textit{Kepler} where the fields overlap. Yields of specific planetary parameter are shown as well as the total.}
    \begin{tabular}{lccccc}
    \hline
    \hline
        Observing Strategy & Including Monos \& Duos & Habitable Zone \& R$_\text{P}$<1.5\,R$_\oplus$ & Period >100\,d & R$_\text{P}$<1\,R$_\oplus$ & Total \\
        \hline
        \textit{PLATO} 2 Years LOPS2 & No & 11$\pm$3 & 421$\pm$19 & 907$\pm$35 & 7766$\pm$87\\
        \textit{PLATO} 4 Years LOPS2 & No & 25$\pm$5 & 1007$\pm$27 & 1897$\pm$45 & 10632$\pm$90\\
        \textit{PLATO} 8 Years LOPS2 & No & 43$\pm$6 & 1556$\pm$39 & 2961$\pm$58 & 12873$\pm$112\\
        \textit{PLATO} 2 Years + \textit{TESS} LOPS2 & No & 12$\pm$4 & 552$\pm$22 & 924$\pm$32 & 7939$\pm$85\\
        \textit{PLATO} 2 Years + \textit{Kepler} LOPN1 & No & 9$\pm$3 & 470$\pm$23 & 945$\pm$33 & 7717$\pm$78\\
        \hline
        \textit{PLATO} 2 Years LOPS2 & Yes & 14$\pm$4 & 782$\pm$26 & 908$\pm$35 & 8127$\pm$90 \\
        \textit{PLATO} 4 Years LOPS2 & Yes & 27$\pm$4 & 1232$\pm$31 & 1897$\pm$45 & 10857$\pm$93\\
        \textit{PLATO} 8 Years LOPS2 & Yes & 44$\pm$6 & 1641$\pm$40 & 2961$\pm$58 & 12959$\pm$112\\
        \textit{PLATO} 2 Years + \textit{TESS} LOPS2 & Yes & 14$\pm$4 & 825$\pm$30 & 925$\pm$32 & 8212$\pm$85\\
        \textit{PLATO} 2 Years + \textit{Kepler} LOPN1 & Yes & 11$\pm$3 & 796$\pm$30 & 945$\pm$33 & 8044$\pm$81\\
    \hline
    \hline
    \end{tabular}
    \label{tab:yields}
\end{table*}

\subsection{Longer Observations of the LOPS2 Field vs. Moving to the Northern Field}
Our computed sensitivities and yields are in agreement with the yields presented in \citet{RedBook_PLATO_2017, Rauer_PLATO_2025, cabrera2026}. Using our sensitivities and yields we compare the two observing strategy scenarios of either continuing a longer pointing in the South or moving to the Northern field and making use of \textit{Kepler} data. We compare these two strategies in terms of new planets that can be detected and were not able to detect to date.

\subsubsection{Longer Observations of the LOPS2 Field}
\begin{figure*}
    \centering
    \includegraphics[width=\linewidth]{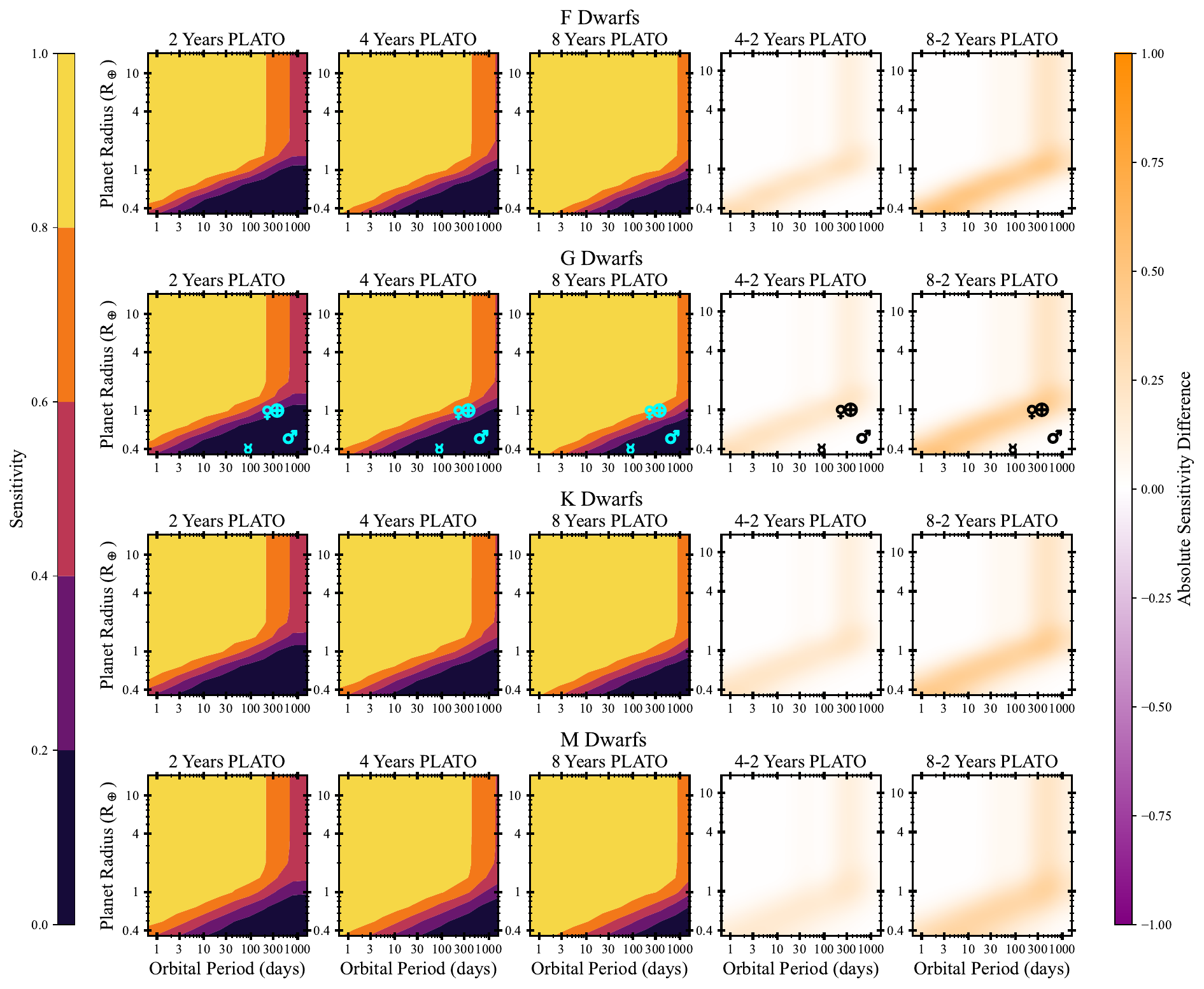}
    \caption{Detection sensitivities on a radius-period diagram of different observing durations of \textit{PLATO}: 2 years , 4 years and 8 years from left to right. The sensitivity difference between of 4 and 8 years of \textit{PLATO} both compared to 2 years of \textit{PLATO} are shown on the right. These are shown for F dwarfs, G dwarfs, K dwarfs and M dwarfs. The Solar System planets Mercury, Venus, Earth and Mars are shown for G Dwarfs in cyan on the sensitivity maps and in black on the difference maps with their respective symbol.}
    \label{fig:PLATO_sensitivities}
\end{figure*}

As \textit{PLATO}'s LOPS2 field would be observed for longer, i.e. 4 or 8 years instead of the planned 2 years, the mission will become more sensitive to detecting small planets and planets of longer orbital period as shown on by the sensitivity and sensitivity difference plots in \autoref{fig:PLATO_sensitivities}. Its ability to detect smaller planets arises from catching more transits of these during a longer observation, resulting in a higher combined signal-to-noise and hence more likely a detection. Observing the field for 4 or 8 years would also allow to catch at least one, if not more, transits of planets with periods beyond 1000 days. As \textit{PLATO} nearly continuously observed the field, only with a short break every 90\,days for rotation and data downlink, the orbital period will be determined from only two transits in most cases. As \textit{PLATO} observes at a short cadence for the Prime sample, the orbital period can be derived well from the transit duration. 

\begin{figure*}
    \centering
    \includegraphics[width=\linewidth]{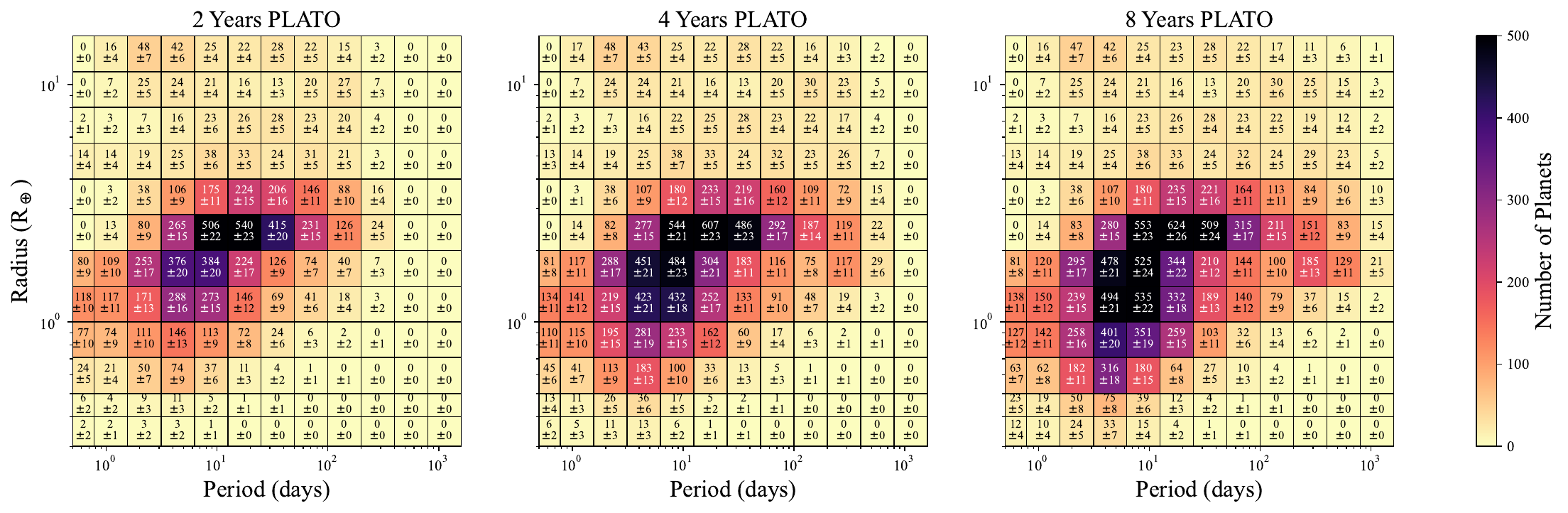}
    \caption{Predicted planet yields in radius-period space accounting for simulated planets that transit at least 3 times. We show the yields for 2 years (left), 4 years (middle) and 8 years (right) of \textit{PLATO} observations of the LOPS2 field.}
    \label{fig:yields_plato}
\end{figure*}

We show the results of our yield estimation in \autoref{fig:yields_plato}. The yields show that \textit{PLATO}'s noise estimations enable to detect many small and long period planets. With the mission continuing to observe the same targets for 4 or 8 years, we find that planets beyond 500\,days become detectable even when requiring at least 3 transits. 
We find that the number of small planets in and out of the habitable zone continues to increase significantly with longer observations as shown in \autoref{tab:yields}. As there are no significant differences between the inclusion of mono- and duo-transits in the detectability of small planets, we see that we require more than 2 transits to detect these small planets and hence a longer baseline of observations is crucial.

We also find that the inclusion on mono- and duo-transits enables to detect more planets beyond 100\,d, especially for the first 2 years of observations. For longer observations this increase stagnates due to our simulations and occurrence rates only considering planets up to 1600\,d. Hence, the vast majority of simulated planets will have transited more than twice during the 8 years of observations. 

Overall, observing the field for up to 8 years opens the possibility to detect an Earth-sized planet at an orbital period of Earth around F, G and K dwarfs. For this case of observing for 8 years, the yields predict that significantly more small and long periods planets will be found, contributing to radius-period space of exoplanets that has been sparsely explored to date due to detection limitations.

\subsubsection{Moving to the Northern Field}
\begin{figure*}
    \centering
    \includegraphics[width=\linewidth]{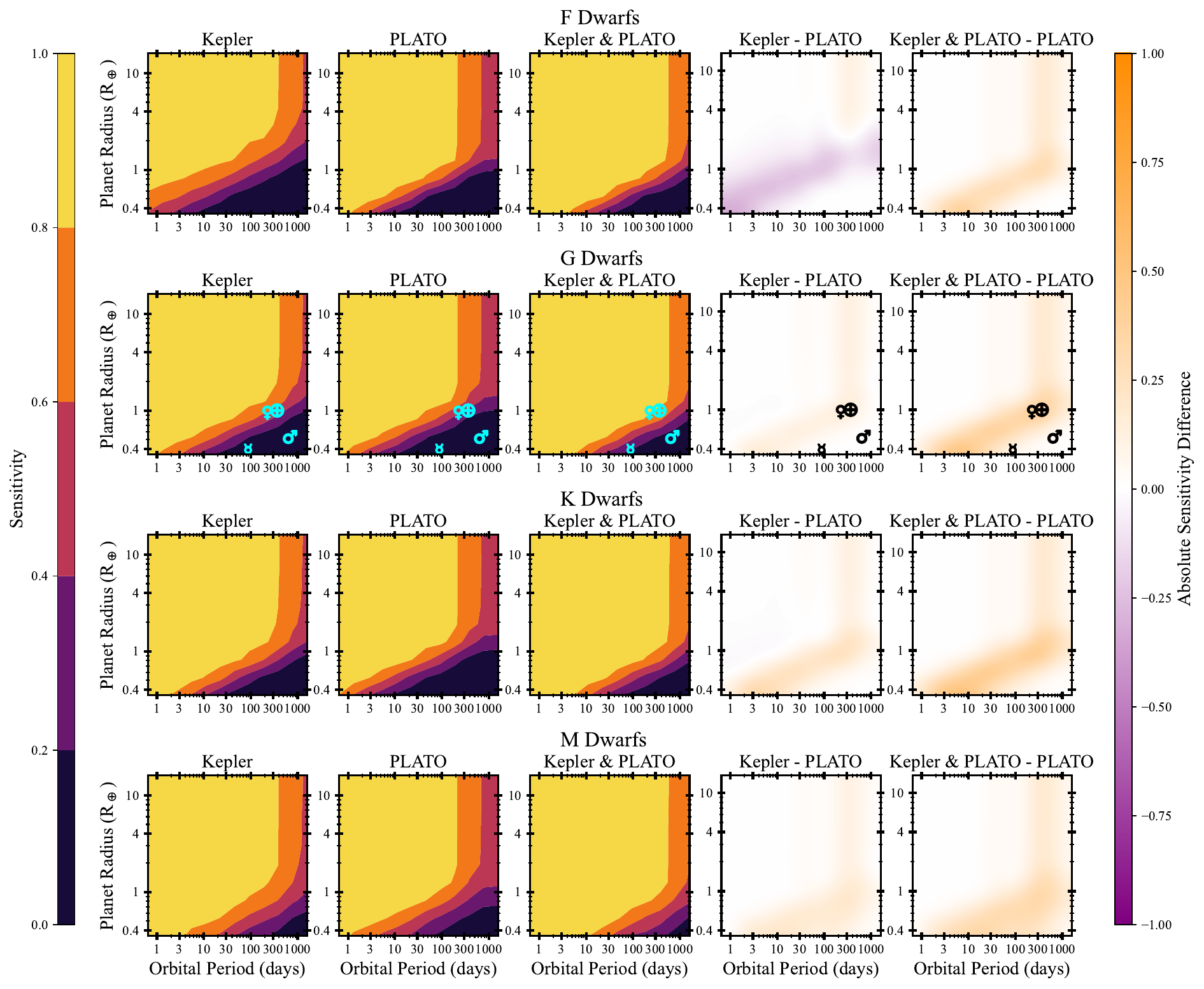}
    \caption{Same as \autoref{fig:PLATO_sensitivities} but comparing \textit{Kepler} only to 2 years of \textit{PLATO} and the combination of \textit{Kepler} and 2 years of \textit{PLATO}.}
    \label{fig:Kepler_sensitivities}
\end{figure*}

Our computed sensitivities for \textit{Kepler} and 2 years of simulated \textit{PLATO} observations and the respective sensitivity differences in \autoref{fig:Kepler_sensitivities} show that \textit{Kepler} is more sensitive to smaller and longer period planets than \textit{PLATO} for G and K dwarfs. However, in terms of F dwarfs, \textit{PLATO} seems to be more sensitive than \textit{Kepler} in detecting smaller planets while \textit{Kepler} remains more sensitive to long period planets. Since the \textit{Kepler} mission observed its field for nearly 4 years and we only simulate 2 years of \textit{PLATO} data, it is apparent that the \textit{Kepler} mission is more sensitive to planets of long periods than the \textit{PLATO} mission. However, as we find varying results based on spectral type in sensitivity in regard to detecting small planets, we analyse the noise properties of the two missions further. 

\begin{figure}
    \centering
    \includegraphics[width=\linewidth]{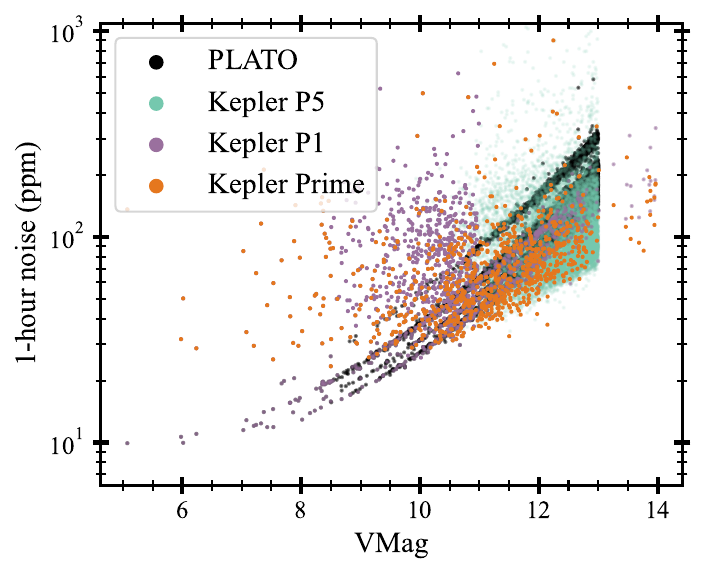}
    \quad
    \includegraphics[width=\linewidth]{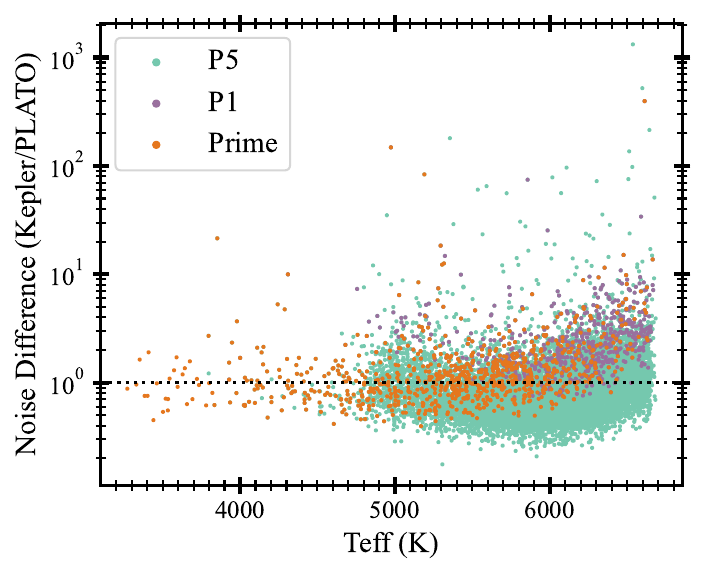}
    \caption{Kepler Noise Properties. Top: Reported 6-hour CDPP scaled to 1 hour of the Kepler targets. We highlight the targets of the Prime sample in orange, P1 in blue and show the P5 sample transparent in the background in green. The predicted 1-hour PLATO noise is shown in purple. Bottom: The ratios between the scaled 1-hour Kepler noise and predicted PLATO noise for the P1, Prime and P5 sample. For reference we show a ratio of one by the black line.}
    \label{fig:kepler_noise}
\end{figure}

As outlined in \citet{Rauer_PLATO_2025}, \textit{Kepler} is reported to reach a better precision for the same star of the same magnitude than the one predicted for \textit{PLATO}. We show the predicted \textit{PLATO}-noise and reported \textit{Kepler} CDPP scaled to 1\,hour for the P1, Prime and P5 sample in \autoref{fig:kepler_noise}. We find that for the fainter \textit{PLATO} stars of the Prime and P5 sample this is the case. However, for the majority of bright stars that make up the majority of the P1 and a fraction of the Prime sample, the observed noise in \textit{Kepler} is higher than the predicted \textit{PLATO} noise. We plot the noise difference between \textit{Kepler} and \textit{PLATO} against the stellar effective temperature for the different samples in \autoref{fig:kepler_noise}. Especially for the bright stars, we find a dependence on temperature hinting at the Kraft-break. We hence conclude that these stars are higher in noise due to stellar activity. As the \textit{Kepler} mission was designed to target planets orbiting fainter stars than the majority of \textit{PLATO} targets, it reaches higher precisions, i.e. lower noise values, for the faint stars of the Prime and P5 sample which is in agreement with \citet{Rauer_PLATO_2025}.
Since F dwarfs make up the smallest and brightest fraction of stars of the Prime sample and are the hottest stars of this, they are mostly impacted by this noise trend seen for magnitude and effective temperature. 

\begin{figure}
    \centering
    \includegraphics[width=\linewidth]{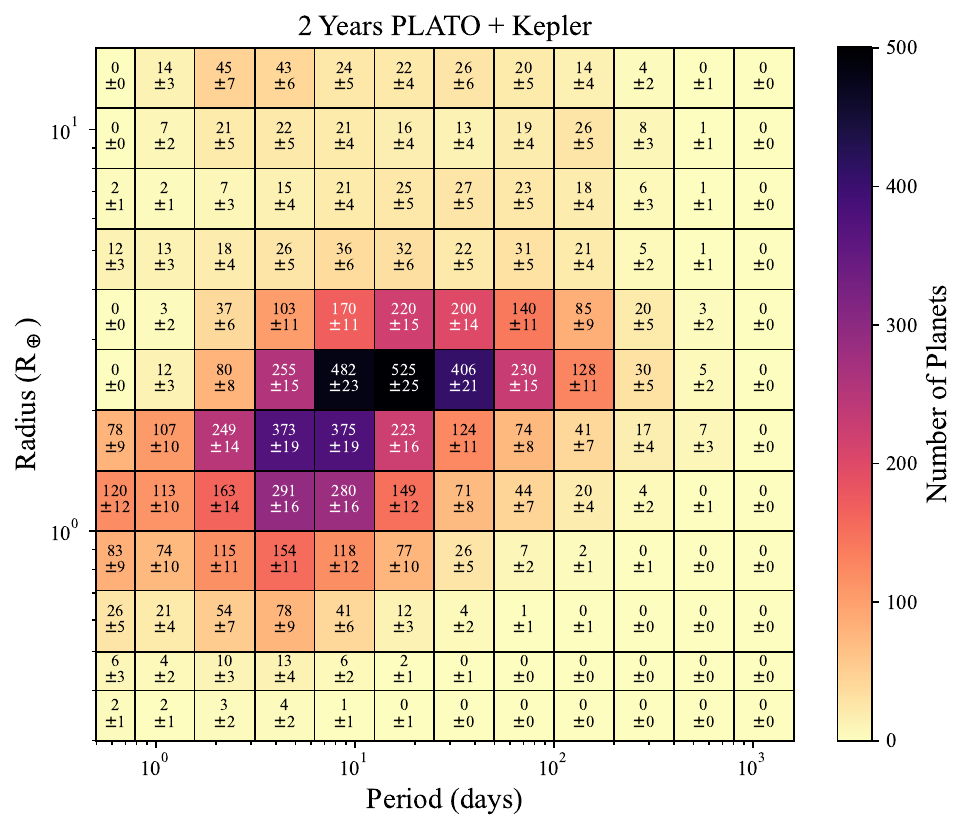}
    \caption{Predicted planet yields for LOPN1 of 2 years of \textit{PLATO} observations, accounting for \textit{Kepler} data where available.}
    \label{fig:yield_kepler}
\end{figure}

We show the results of our yield computation of planets with at least 3 transits for stars of the LOPN1, accounting for \textit{Kepler} data where it is available, in \autoref{fig:yield_kepler}. We note that the sample of LOPN1 is slightly smaller than the sample of LOPS2 but similar enough to compare these yields to those of 2 years of \textit{PLATO} observations in LOPS2 which are shown in \autoref{fig:yields_plato}.  The combination of \textit{PLATO} and \textit{Kepler} data enables to detect planets of longer orbital periods than \textit{PLATO} only. However, as the overlap of the 2 missions only covers a small fraction of stars, these additional detections to no significantly contribute to the yields (see \autoref{tab:yields}). We find that the total yield of observing the LOPS2 field and the LOPN1 field for 2 years each result in a higher total yield of planets but a smaller yield for smaller, long period planets compared to 4 years continuous observations of LOPS2.


Overall, we find that the combination of \textit{Kepler} and \textit{PLATO} data enables us to be sensitive to planets as small as Earth and Venus on a similar orbital period for F, G, K and M dwarfs. However, we note that this is only for a small sample of stars as the \textit{Kepler} field only makes up a fraction of LOPN1. With this the combination of the two missions does not significantly contribute to the overall yields.

\subsection{The Impact of \textit{TESS} data}
We show the \textit{TESS} and \textit{PLATO} sensitivities individually in \autoref{fig:tess_sensitivities} and \autoref{fig:tess_sensitivities_subgiants} updated for the Prime sample. As found in \citet{Eschen_PLATO_Sensitivities_2024} and expected from the precision estimations of each mission \citep{Rauer_PLATO_2025}, we find that the \textit{TESS} sensitivity increases towards smaller and longer period planets with more data available, i.e. \textit{TESS} is more sensitive to smaller and longer period planets around targets that lie in the CVZ. As the \textit{PLATO} mission provides long term monitoring at a higher precision, its significantly more sensitive to smaller and longer period planets than \textit{TESS}, even in comparison to targets in the CVZ. 

\begin{figure*}
    \centering
    \includegraphics[width=\linewidth]{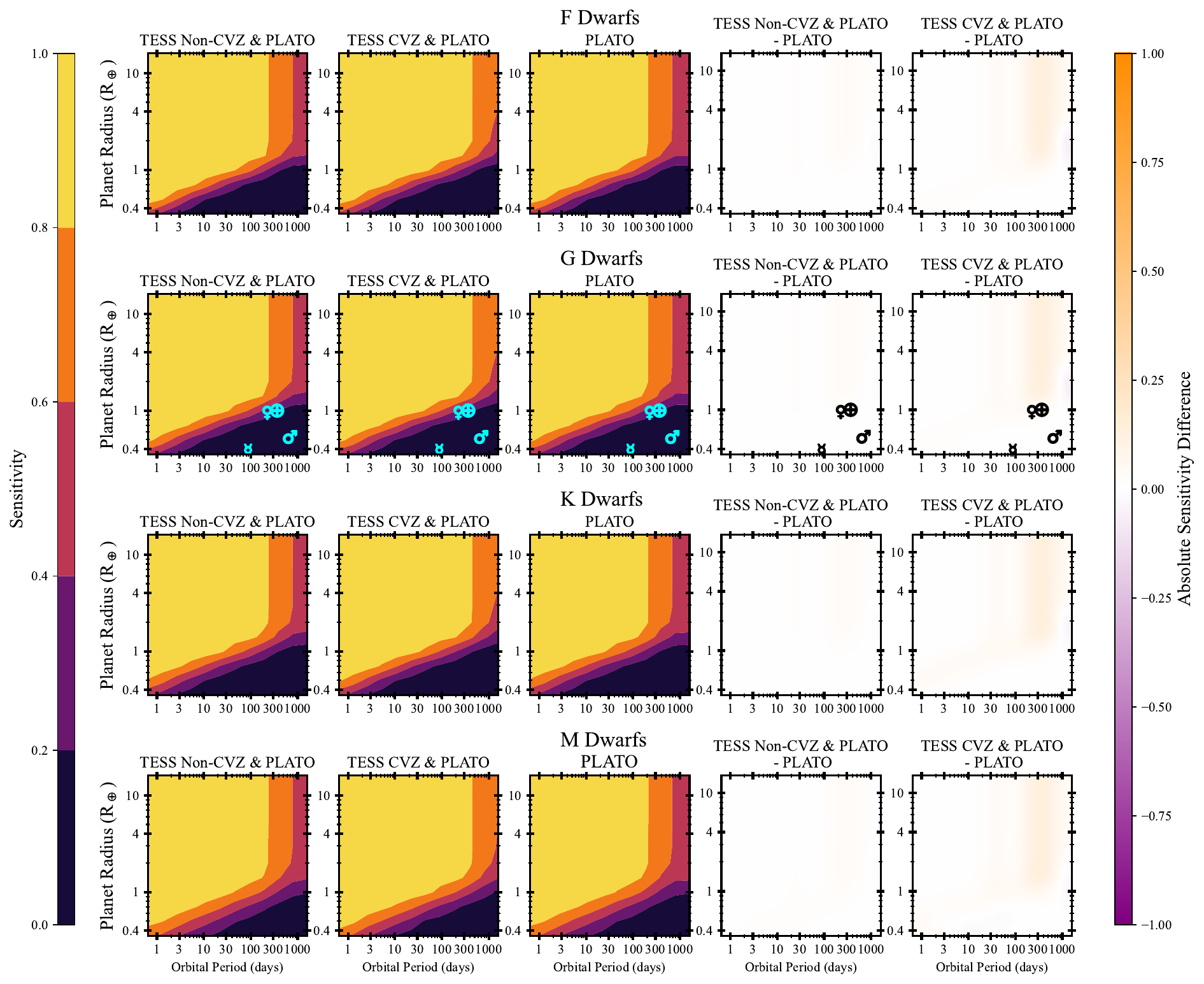}
    \caption{Same as \autoref{fig:PLATO_sensitivities} but comparing the combination of \textit{TESS} and 2 years of \textit{PLATO}, differentiating between targets in the CVZ and non-CVZ, to only 2 years of \textit{PLATO}.}
    \label{fig:tess_plato_sensitivities}
\end{figure*}

We show the sensitivities for the combination of \textit{TESS} and \textit{PLATO} data differentiated by targets that lie in the CVZ and Non-CVZ in comparison to 2 years of \textit{PLATO} observations in \autoref{fig:tess_plato_sensitivities} and \autoref{fig:tess_plato_sensitivities_subgiants}. As \textit{PLATO} is predicted to reach higher precisions than \textit{TESS}, adding the \textit{TESS} data to the \textit{PLATO} data does not significantly impact the detection sensitivities towards smaller planets. Targets outside of the CVZ have not many \textit{TESS} sectors available and hence only marginally improve the sensitivities. However, targets in the CVZ have more than 30 sectors of \textit{TESS} data and with this improve \textit{PLATO}'s sensitivities in period. Since the combination of \textit{TESS} and \textit{PLATO} data provides a longer baseline than only 2 years of \textit{PLATO} observations, the combination of the two improves the detection sensitivity up to 1000\,d. As targets in the CVZ have been monitored continuously every other year, this more than doubles the baseline and hence enables the detection of planets of longer orbital periods. The \textit{TESS} mission also continues its observations, including the LOPS2 field, which will increase this impact, especially in longer periods as more data will become available in the future.

Overall, we find that the addition of \textit{TESS} data does not significantly improve \textit{PLATO}'s detection sensitivities on average. However, targets in the CVZ provide a longer baseline and hence can contribute to detecting planets of longer orbital periods. Furthermore, even small increases in sensitivity can change a non-detection to a detection. As our sensitivities are averaged these impacts can be larger or smaller for individual systems and should be accounted for. Hence including the available \textit{TESS} data can have an impact on the detectability of planets, especially of long periods.

\begin{figure}
    \centering
    \includegraphics[width=\linewidth]{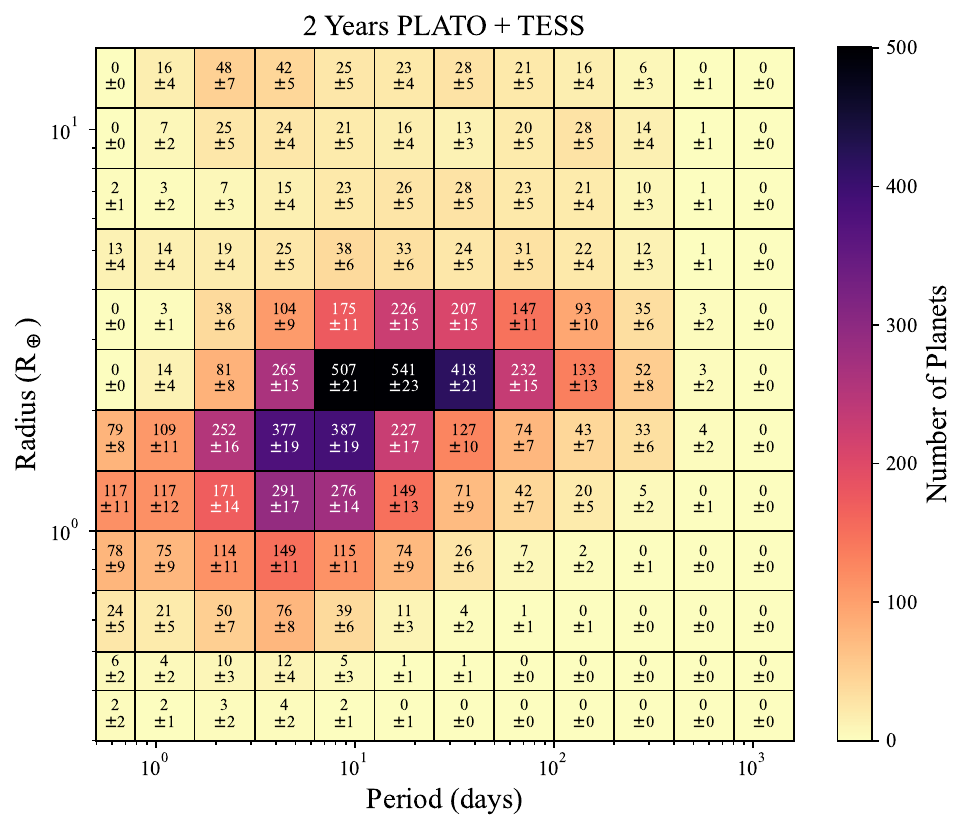}
    \caption{Predicted planets yields for 2 years of \textit{PLATO} observations combined with \textit{TESS} data.}
    \label{fig:yields_tess}
\end{figure}

\section{Discussion}
\label{sec:discussion}
As the \textit{PLATO} mission not only follows the goal to detect more small planets of long orbital periods, it also aims to characterise these well through ground-based follow-up and ultimately derive occurrence rates for these systems that were challenging to detect to date. Hence, when comparing the observing strategies, these aspects have to be taken into account. We will hence discuss the impact of each observing strategy to the ground-based follow-up and the planet output.
As we previously discussed for \textit{Kepler} we will also include a discussion on stellar activity.

\subsection{Assessment of Potential New Detections}

Our computed sensitivities and yields show the (dis-)advantages of each observing strategy. We find that the monitoring of the LOPS2 field for the first 2 years of the \textit{PLATO} mission provides a greater sensitivity to smaller and longer period planets to \textit{TESS} and will contribute to the growing sample of Super-Earths and Sub-Neptunes of varying orbital periods and hence equilibrium temperatures. However, monitoring the field for only 2 years, does not provide the detection sensitivity for a real Sun-Earth analogue, nor does it allow to detect planets at periods beyond $\sim$350\,d if three or more transits are required for a detection. Such systems become detectable according to our sensitivities and yields through longer monitoring of the LOPS2 field. To become sensitive to detecting a real Sun-Earth analogue, we find the highest detection sensitivity when monitoring the field for up to 8 years continuously. This scenario would also enable to detect transiting planets of orbital periods beyond 1000\,d.

In the scenario of \textit{PLATO} moving to the North after two years in the South, the total detections would increase as more stars are monitored. However, the sensitivity and yields towards a Sun-Earth analogue or planets above 1000\,d would remain low. Even the combination of the \textit{PLATO} and \textit{Kepler} data, would not yield many new systems as the overlap of the two missions is small and the majority of the small and long period planets in the overlap have already been found by \textit{Kepler} due to its higher precision and continuous monitoring for four years. 

Hence, monitoring both fields will enable to detect the highest number of planets due to the monitoring of more stars, but continuously monitoring the Southern field for longer will increase the detection sensitivity towards smaller and longer period planets opening the space to be sensitive to the detection of real Sun-Earth analogues and planets beyond 1000\,d which will push into a regime of planets that has been sparsely explored to date. 

\subsection{Occurrence Rates}
Beyond affecting the number of new planet detections, the different observing strategies will also affect \textit{PLATO}'s impact on occurrence rates. Monitoring both fields would provide the largest overall sample and therefore improve the statistical precision with which demographic features at shorter periods can be measured. In particular, the increased number of small planets would help refine features such as the radius valley \citep{OwenWu2013, Fulton2017} and hot Neptune desert \citep{Mazeh2016,Lundkvist2016}, including their locations, widths, and statistical significance as a function of stellar properties. This larger sample could also enable more detailed comparisons of planet populations around FGK and M dwarfs, where current measurements from \textit{TESS} suggest that the structure of the small planet population may differ between stellar types \cite[e.g.,][]{Cui2026,Gillis2026}.

In contrast, continuously monitoring LOPS2 for several years would provide fewer total stars but substantially greater sensitivity to long-period planets. Our simulations show that \textit{PLATO} would be able extend transit-based occurrence rate measurements beyond the $\lesssim$400\,d regime over which Kepler provides its strongest constraints, and eventually into periods of $\gtrsim1000$\,d. Such measurements would provide an important complement to occurrence rates inferred from radial velocity surveys, which currently dominate our knowledge of giant planet populations at these orbital periods. Importantly, \textit{PLATO} could probe this regime using transits while simultaneously detecting the shorter-period super-Earths and sub-Neptunes in the same systems. This would enable more self-consistent tests of correlations between outer giant planets and inner small planet populations, without relying on populations drawn from fundamentally different detection methods as is currently required \cite[e.g.,][]{BryanLee2024,Chen2026}.

In both cases, converting the resulting planet samples into robust occurrence rates will require a uniform automated search and vetting procedure, empirical injection-recovery measurements of completeness, and a characterization of false-alarm and astrophysical false-positive rates. These selection effects will be especially important when comparing occurrence rates across stellar populations and observing strategies, since the sensitivity depends strongly on stellar properties, observing baseline, and the number of transits observed.

The overlap with \textit{TESS}, and in the Northern field with \textit{Kepler}, provides an additional opportunity for occurrence rate studies. Combining data sets can extend the time baseline and improve sensitivity to long-period planets, particularly for stars in the \textit{TESS} continuous viewing zones. However, the heterogeneous observing histories of these stars will need to be incorporated into any demographic analysis.

\subsection{Ground-based Follow-Up}

In addition to detecting new planets and measuring their radii precisely through the transit method, a mission-critical part of \textit{PLATO} is also measuring the masses of these newly found planets precisely using ground-based spectroscopic instruments, such as ESPRESSO \citep{Pepe_ESPRESSO_2021} and HARPS/NIRPS \citep{Pepe_HARPS_2000,Mayor_HARPS_2003,Bouchy2022} \citep{Rauer_PLATO_2025}. Obtaining accurate masses through RV observations requires deep understanding of both planetary systems (host stars and orbiting bodies) and utilised instrumentation. To this end, the {\em PLATO} ground-based follow-up is naturally required to collect RV observations of exciting Earth-twin systems for decades. This will enable assessment of RV instrument stability and stellar activity processes that can hinder planet characterisation (see below). Stellar activity mitigation in RV data is thought to be aided by combining contemporaneous spectroscopic and photometric exoplanetary data \citep{Barragan2026}, however this needs further study \citep{Cretignier2026}. Beyond magnetic activity cycles, continuous photometric monitoring of stellar granulation signals could aid modelling of such processes in RV data as granulation and super-granulation can be a hindrance to exoplanet mass measurements \citep{Lakeland2024,OSullivan2025}. Whereas, multiple transits of orbiting bodies also helps ephemerides refinement that aids RV modelling. Thus, if {\it PLATO} continue LOPS2 observations for longer than 2 years, RV mass modelling of discovered planets, including long-period objects, would be improved. Conversely, if {\it PLATO} moves to a Northern field more overall systems could be monitored by RV instruments, especially those in the Northern hemisphere (i.e. HARPS-N \citep{Cosentino_HARPSN_2012}, HARPS3 \citep{Thompson2016}, etc.). However, support from continuous {\em PLATO} observations would be limited.

\subsection{The Impact of Stellar Activity}

Stellar interior and surface processes (see \citealt{Jeffers2023} and \citealt{Bellotti2025} for reviews) are photometrically and spectroscopically observable, and alter exoplanet light curve and RV time series (e.g. \citealt{Haywood_Corot7_2014,Damasso_Young_Planetary_System_HARPSN_2020}). Pressure-mode oscillations in Sun-like stars cause photometric variations on minute timescales whose power spectral density analysis underpins asteroseismological determination of stellar physical parameters (e.g. \citealt{Chaplin2013}). Such periodic variations can have photometric amplitudes of 10-100s\,ppm with this signal significantly reducing when light curve data is binned to timescales longer than 15 minutes. Near-surface convective processes in Solar-like stars produce granulation features whose vertical movement of material is detectable in photometric data. This typically yields variations on the order of 0.01\% with timescales of minutes to hours for FGK-dwarfs \citep{Kallinger2014}. Similarly to oscillations, activity from granulation can typically be averaged over, however care must be taken for photometric exoplanet detection as granulation signals can still remain correlated on ingresses/egresses timescales \citep{Morris2020,Sulis2020,Krenn2024}. Moreover, flares and superflares can contribute to significant flux increases over minutes to hours timescales \citep{Balona2015}. However, for mature (>1\,Gyr) FGK-dwarfs, such events are not common \citep{Namekata2017} and so should not impact transit detection on a population level.

Magnetically-driven active surface features, such as spots, plages, and faculae, can result in significant photometric variability on the timescales of stellar rotation periods (days to months) due to temperature, and thus contrast, differences between these regions and the quiet stellar surface \citep{Bruno2023}. Star spots in particular can induce variations of 100\,ppm to 0.5\% for mature stars, and up to 10\% for younger stars. Importantly, the amplitude strength also varies over year to decade timescale magnetic cycles. These substantial effects can make transit detection of exoplanet severely challenging, requiring dedicated analysis pipelines \citep{Battley2020}. Moreover, the strength of all aforementioned signals are often unknown prior to observations and may alter between seasons. Therefore, noise estimates of individual stars may be underestimated. However, this does not effect population level analysis and conclusions.

\subsection{TTVs}

Planet-planet gravitational interactions modify the orbits of transiting exoplanets, inducing transit timing variations \citep[TTVs][]{Agol2005}. The amplitude and timescale of TTVs greatly depend on the orbital configurations and planetary masses. Over the four years of baseline of the Kepler mission, significant TTVs were measured for hundreds of KOIs \citep[10 to 60\% of the 2599 analysed KOIs][]{Holczer2016,Yahalomi2023}.

Modelling these TTVs enables constraints on the orbital parameters and planetary masses of the interacting planets. Systems near mean motion resonances (MMR), where the ratio of orbital periods is close to a fraction of small integers, are particularly suited to the characterization by TTVs. Indeed, these configurations typically generate TTVs of tens of minutes to hours of amplitudes, over timescales of months to a few years. For a planet near but not into MMR, the TTVs are typically faster, but the mass determination can be subject to mass-eccentricity degeneracy. It requires good precision on the planet timings, or RV follow-up, to break this degeneracy \citep{Boue2012,Leleu2023}. For a pair of planets inside a MMR, the mass and (part of the) eccentricities can be measured as long as a full period of the TTVs is observed. In this case, the TTV period is proportional to the planets' orbital period, and $\sim ((m_1+m_2)/m_\star)^{-2/3}$ \citep{Nesvorny2016}. Longer observational baseline therefore enables the mass determination of a planet at a longer orbital period, and/or less massive. The characterization of a resonant pair of super-Earth at 10 days of orbital period requires a $\sim 4 $ year baseline \citep[e.g. ][]{Leleu2021}.

Alternatively, the effect of planet-planet conjunction, also called 'chopping' can be observed \citep{Deck2014}. This requires the period ratio between planets to be small ($\lesssim 2$) but not resonant. The amplitude of this signal is typically smaller than the resonant one, but it can be observed with only a few transits. It is also proportional to the mass of the perturber and the planet's orbital period, making it ideal for the characterization of long-period planets. Earth-Venus analogue around a sun-like star would produce 4-min TTVs. This is comparable to the expected PLATO precision for the transit of an Earth around a Sun-like star of V=10\,mag with 24 cameras. The signal would be boosted if the pair had a smaller period ratio, or would be easier to measure if the Earth was paired with a larger planet. Here, a longer baseline helps accumulating more signal, and can also enable diagnosing the presence of unseen perturber. Finally, the brightness of PLATO's targets should enable a combination of TTV and RV measurements, in particular in the case of non-transiting perturbers. 


Observing TTVs of known transiting exoplanets with ultra short periods ($P \lesssim 2$\,days) can also contribute to the study of tidal processes in these systems. Monitoring these systems over multi-decade timescales can reveal the gradual decrease in orbital period resulting from this process and measuring this can allow for constraints on the tidal quality factor of the host star, a key parameter for quantifying tidal dissipation \citep{Goldreich1966,Barker2020}.

Within the LOPS2 field NGTS-6\,b \citep{Vines2019}, NGTS-10\,b \citep{McCormac2020} and HATS-70\,b \citep{Zhou2019b} are likely to exhibit orbital decay. However, studies to date, have not been found evidence yet \citep[][ Romanec et al. (in prep.)]{Griffiths2026}, hence long-term monitoring of these systems and extending the baseline with \textit{PLATO}'s high precision photometry can contribute to these and upcoming detections. 

While continuous photometric observations for orbital decay systems is expected to enable precise determination of the ephemerides, the robust detection of decay, significantly benefits from transit observations taken at periodic intervals over two or more decades as would be the case when combining \textit{Kepler} and \textit{PLATO} data in the North as shown by \citet{Mitchell_PLATO_TTVs_2026} or using the available \textit{TESS} data in combination with long observations of the LOPS2 field.

Hence, \textit{PLATO}'s photometric precision but also long-term monitoring in combination with the \textit{Kepler} and/or \textit{TESS} mission will provide precise ephemerides and enable more robust detections of orbital decay in ultra-short period systems.

\section{Conclusions}
\label{sec:conclusions}

We have assessed \textit{PLATO}'s observing strategies for the two scenarios of either staying in the South for longer or moving to the North after two years of observations in the South in terms of sensitivities and yields. 

In terms of sensitivities, we find that monitoring a field, in this case the LOPS2 field, for a longer duration increases \textit{PLATO}'s sensitivity towards detecting Sun-Earth analogues and long period planets. With this, \textit{PLATO} would push planet discoveries into a radius-period space that is little explored by current and previous planet-hunting missions. 
However, moving to a Northern field would increase the number of stars monitored and hence the total predicted planet yields of the mission. The overlapping region with \textit{Kepler} has been mainly explored by \textit{Kepler}'s higher precision and longer monitoring that \textit{PLATO} does only marginally contribute to new detections around the overlapping stars that however only make up a fraction of the LOPN1 field. 

We discussed the impact the two observing strategies on the ground-based follow-up to determine the planets' masses
and the detection on TTV and Orbital Decay signals. We find that the characterisation of the newly detected targets will be more feasible in the South due to high-precision ESO instruments such as ESPRESSO. A longer baseline will also enable to characterise and gain deeper insights into TTV and Orbital Decay systems, which can be achieved by a longer monitoring of the South but also by combining \textit{PLATO} and \textit{Kepler} monitoring. The later however, only applies to a small fraction of targets. 

Overall, we find that each observing strategy has its own benefits and monitoring both fields allows to detect the highest number of planets, but staying in the Southern field for longer enables the detection and characterisation of the most exciting, to date unexplored, systems of small and long period planets.

\section*{Acknowledgements}
We thank Roland Vanderspek for discussions on TESS.
This work presents results from the European Space Agency (ESA) space mission Plato. The Plato payload, the Plato Ground Segment and Plato data processing are joint developments of ESA and the Plato Mission Consortium (PMC). Funding for the PMC is provided at national levels, in particular by countries participating in the Plato Multilateral Agreement (Austria, Belgium, Czech Republic, Denmark, France, Germany, Italy, Netherlands, Portugal, Spain, Sweden, Switzerland, Norway, and United Kingdom) and institutions from Brazil. Members of the Plato Consortium can be found at \url{https://platomission.com/}. The ESA Plato mission website is \url{https://www.cosmos.esa.int/plato}.
This paper made use of data collected by the TESS mission and are publicly available from the Mikulski Archive for Space Telescopes (MAST) operated by the Space Telescope Science Institute (STScI). Funding for the TESS mission is provided by NASA’s Science Mission Directorate. We acknowledge the use of public TESS data from pipelines at the TESS Science Office and at the TESS Science Processing Operations Center. Resources supporting this work were provided by the NASA High-End Computing (HEC) Program through the NASA Advanced Supercomputing (NAS) Division at Ames Research Center for the production of the SPOC data products.
YNEE acknowledges support from a Science and Technology Facilities Council (STFC) studentship, grant number ST/Y509693/1. 
TGW acknowledges support from the UKSA and the University of Warwick. 



\bibliographystyle{mnras}
\bibliography{example} 




\appendix

\section{\textit{TESS} Coverage of \textit{PLATO} targets}
\begin{figure}
    \centering
    \includegraphics[width=\linewidth]{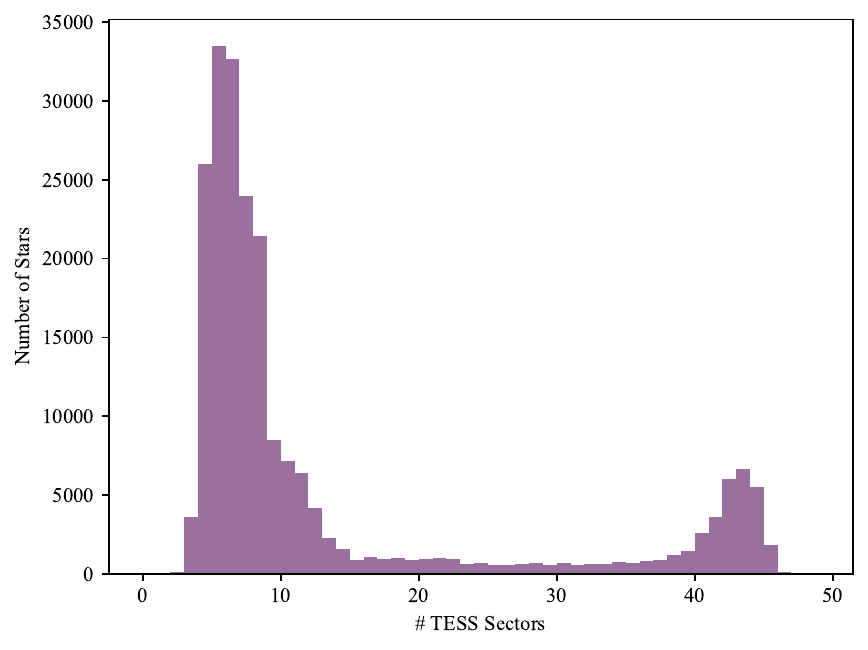}
    \caption{Histogram showing the number of \textit{TESS} sectors each PIC target in the LOPS2 field has been observed in.}
    \label{fig:tess_sectors}
\end{figure}

\section{Estimating the \textit{TESS} noise}
\label{app:tess_noise}

While all targets in the tPIC and Prime sample have \textit{TESS} data, not all have been processed by SPOC. We use all Prime targets with available SPOC data to estimate the noise. 

Lightcurves processed by the SPOC pipeline report the Combined Differential Photometric Precision \citep[CDPP;][]{Jenkins_Kepler_Pipeline_2010,christiansen_cdpp_2012} for 30\,min, 1\,hr and 2\,hr. However, \textit{TESS} is not reporting the CDPP-noise in sectors 40-68. Hence, we computed the 2-hour-noise using the \textsc{lightkurve} \texttt{estimate\_cdpp} function which is an implementation of the methods described in \citet{Gilland_KeplerNoise_2011}. To compute the transit duration in the given cadence, we use a value of 0.25 up to sector 26, 0.083 for sectors 27 to 55 and a value of 0.028 for sectors 56 onwards since we are using the long cadence FFI data. Following \citet{Gilland_KeplerNoise_2011}, we use the  Savitsky-Golay filter fitting a 2nd order polynomial over a 2\,day window width and remove outliers above 5$\sigma$. 

\begin{figure}
    \centering
    \includegraphics[width=\linewidth]{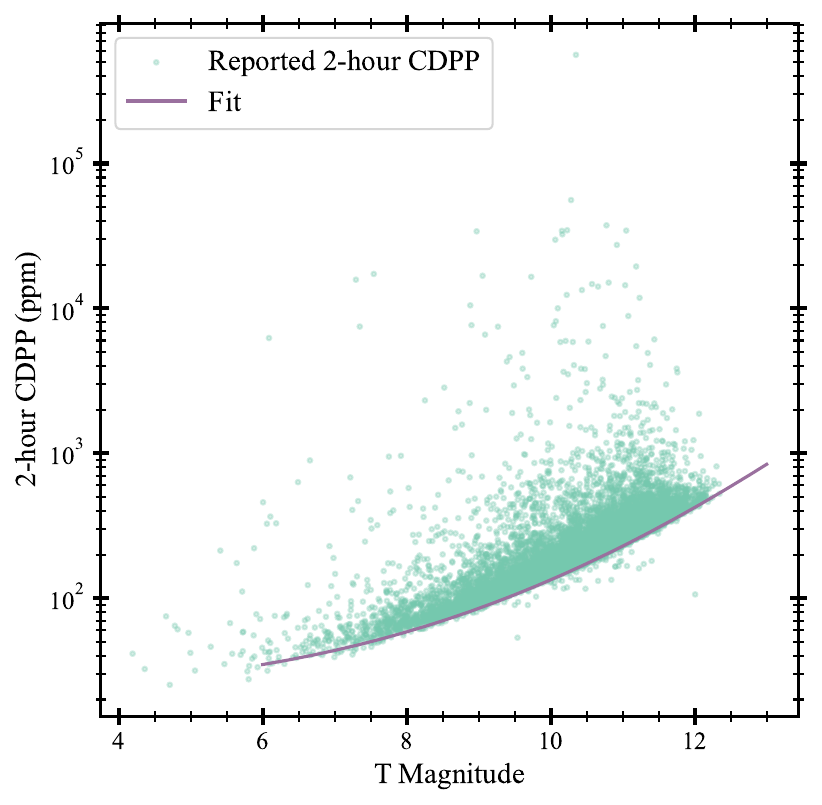}
    \caption{Averaged 2-hour CDPP from the reported and computed values of each Prime targets against their T magnitude are shown in green. The function fitting the lower boundary is shown in purple.}
    \label{fig:tess_noise}
\end{figure}

Using the average CDPP for each star and equation 1 of \citet{Kunimoto_TESS_Yield_2022}, we fit the lower boundary of the 2-hour-CDPP as done in \citet{Eschen_PLATO_Sensitivities_2024}. We show the CDPP distribution and our best fit function of the lower boundary in \autoref{fig:tess_noise}.

\section{\textit{TESS} Sensitivities}
\begin{figure*}
    \centering
    \includegraphics[width=\linewidth]{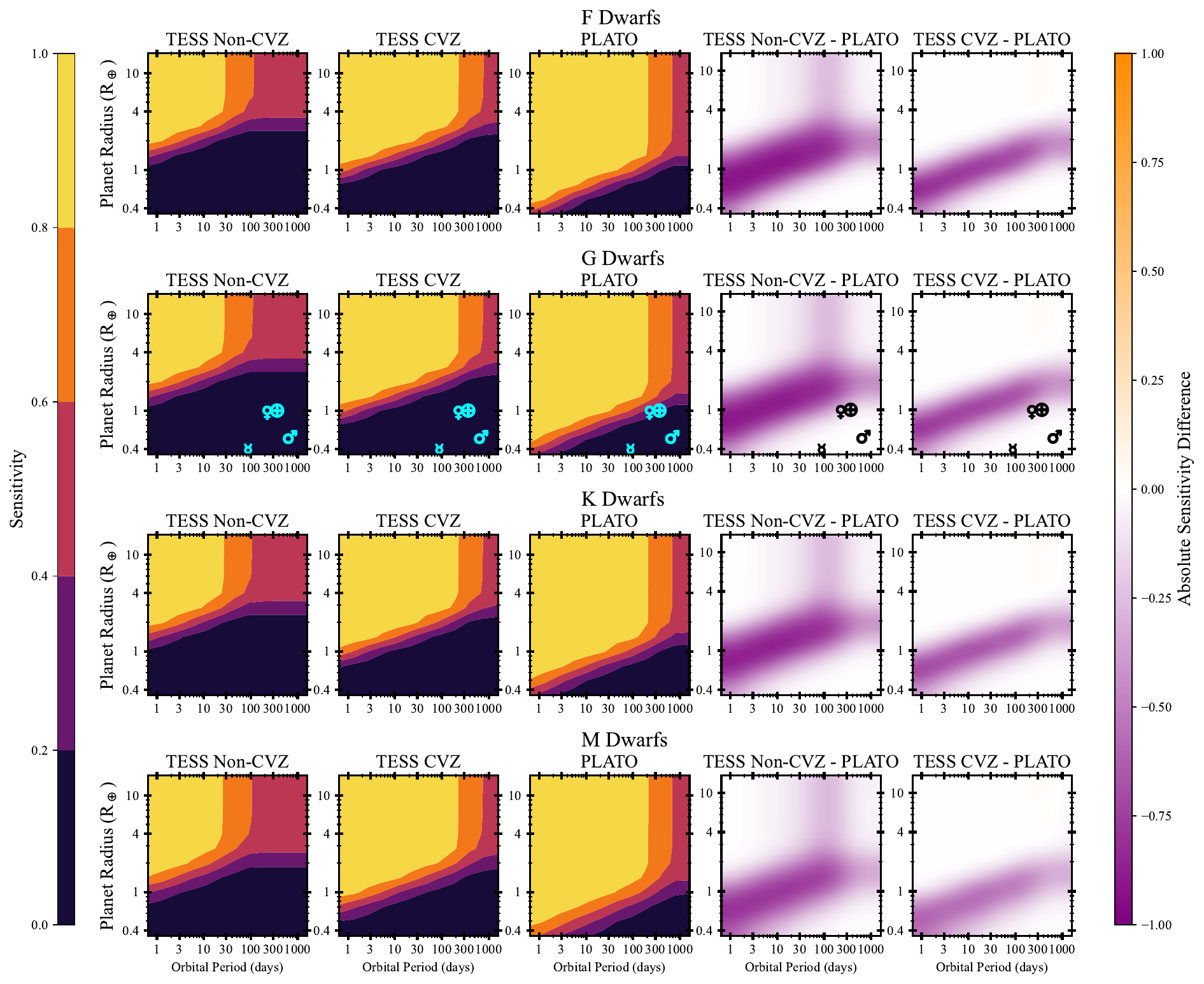}
    \caption{Same as \autoref{fig:PLATO_sensitivities} but comparing \textit{TESS}, differentiating between targets in the CVZ and non-CVZ, to 2 years of \textit{PLATO}.}
    \label{fig:tess_sensitivities}
\end{figure*}

\section{Sensitivities for sub-giants}
\begin{figure*}
    \centering
    \includegraphics[width=\linewidth]{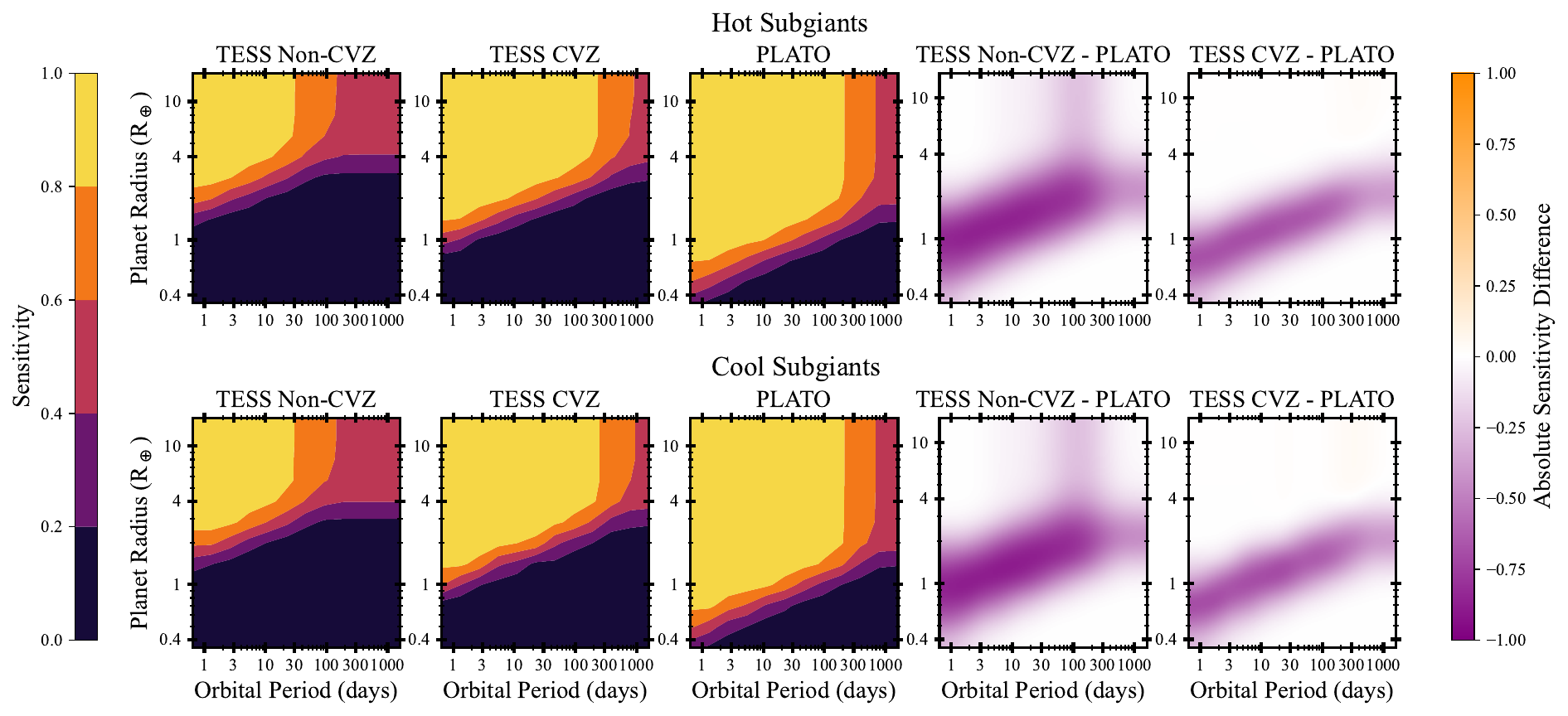}
    \caption{Same as \autoref{fig:tess_sensitivities}, but for subgiants.}
    \label{fig:tess_sensitivities_subgiants}
\end{figure*}
\begin{figure*}
    \centering
    \includegraphics[width=\linewidth]{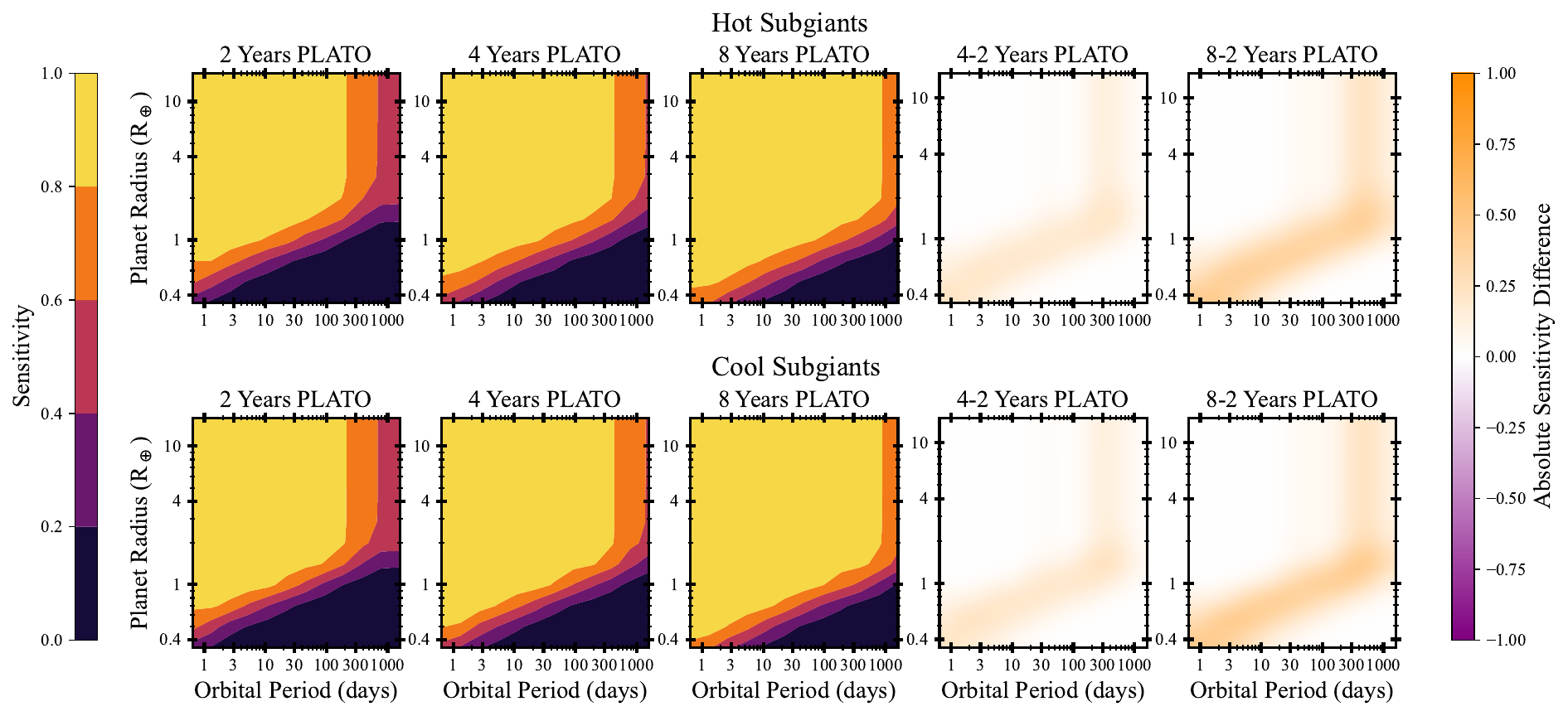}
    \caption{Same as \autoref{fig:PLATO_sensitivities}, but for subgiants.}
    \label{fig:plato_sensitivities_subgiants}
\end{figure*}
\begin{figure*}
    \centering
    \includegraphics[width=\linewidth]{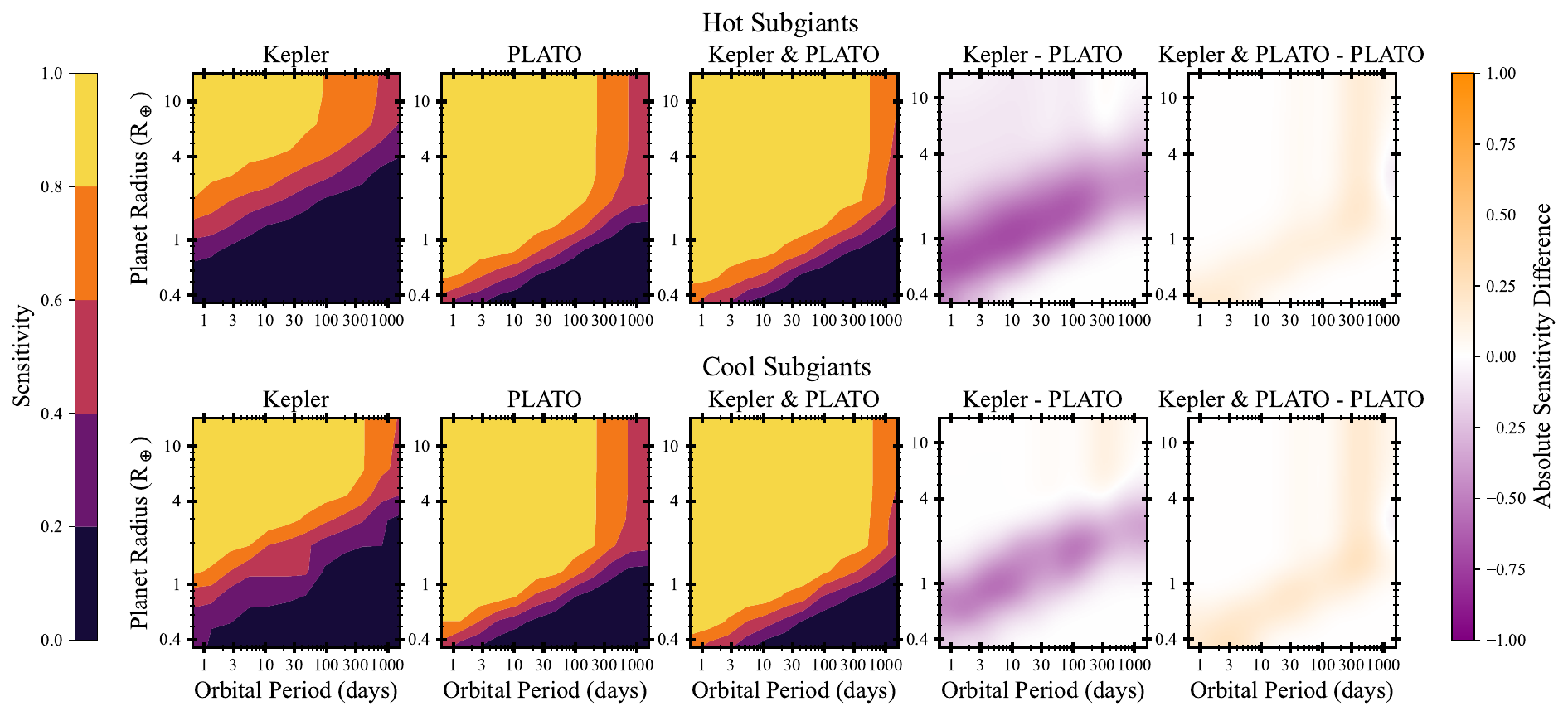}
    \caption{Same as \autoref{fig:Kepler_sensitivities}, but for subgiants.}
    \label{fig:kepler_sensitivities_subgiants}
\end{figure*}
\begin{figure*}
    \centering
    \includegraphics[width=\linewidth]{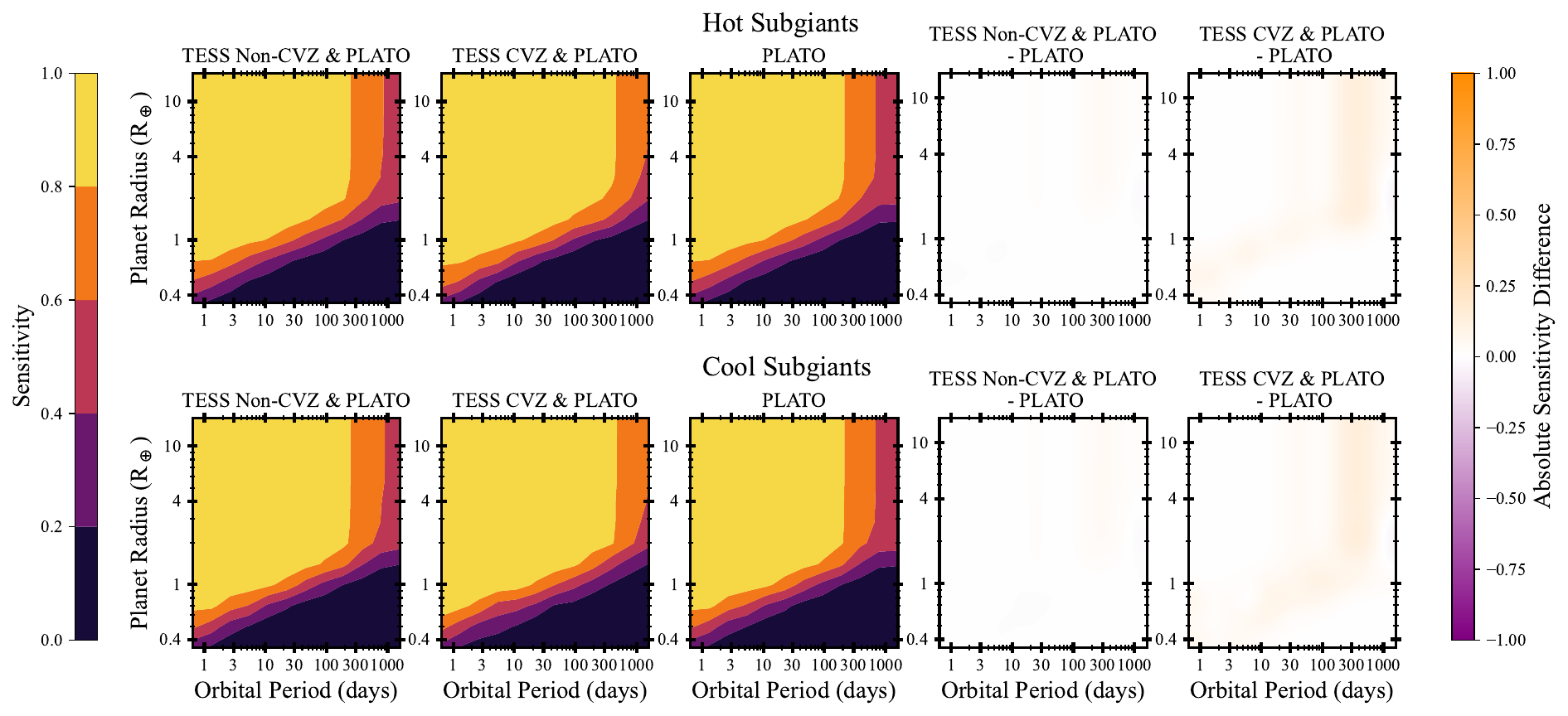}
    \caption{Same as \autoref{fig:tess_plato_sensitivities}, but for subgiants.}
    \label{fig:tess_plato_sensitivities_subgiants}
\end{figure*}

\section{Extrapolated Occurrence Rates}
\begin{figure}
    \centering
    \includegraphics[width=\linewidth]{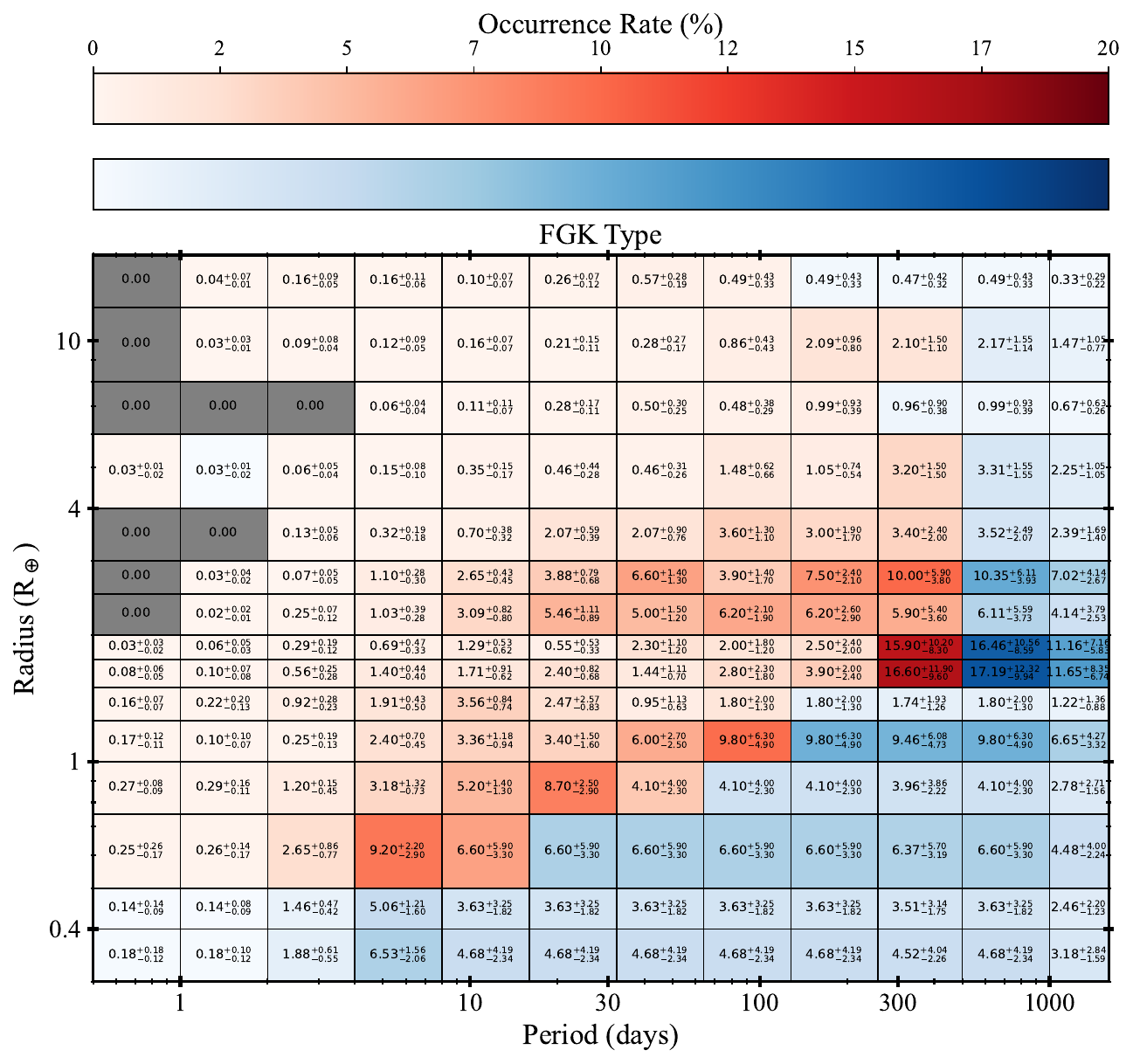}
    \caption{Extrapolated Occurrence Rates from \citet{Hsu_OccurrenceRates_2019} for F, G and K type stars. Occurrence rates with reported values are shown by the red colormap, extrapolated occurrence rates by the blue colormap and reported upper limits that are lower than extrapolated occurrence rates are set to 0 and shown in gray.}
    \label{fig:occurrence_rates_fgk}
\end{figure}

\begin{figure}
    \centering
    \includegraphics[width=\linewidth]{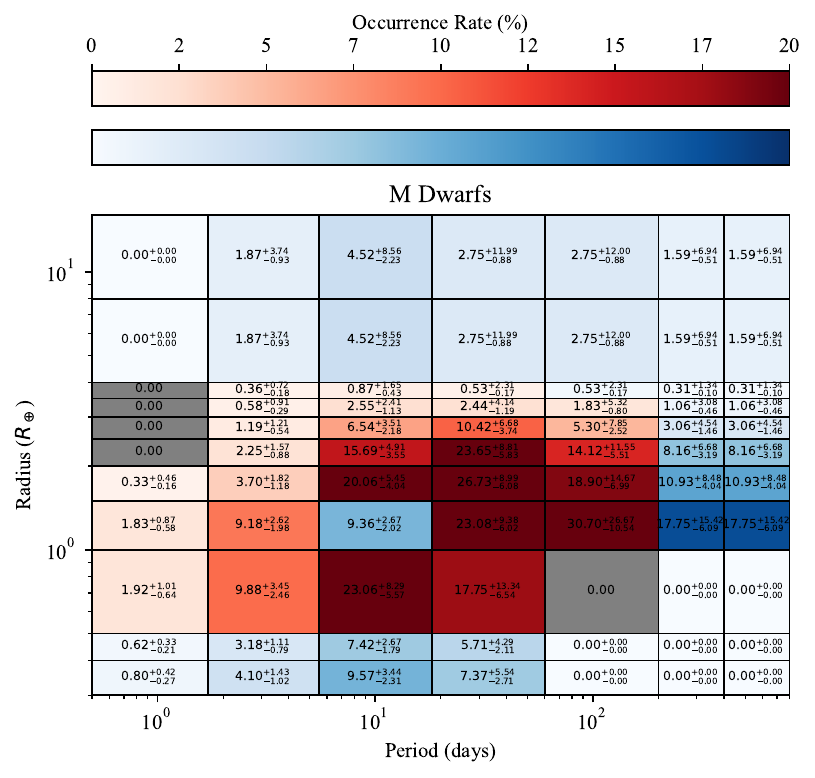}
    \caption{Extrapolated Occurrence Rates from \citet{Dressing_Charbonneau_Occurrence_2015} for M Dwarfs. Occurrence rates with reported values are shown by the red colormap, extrapolated occurrence rates by the blue colormap and reported upper limits that are lower than extrapolated occurrence rates are set to 0 and shown in gray.}
    \label{fig:occurrence_rates_m}
\end{figure}


\bsp	
\label{lastpage}
\end{document}